\documentclass[prd,floatfix,superscriptaddress,nofootinbib,nobibnotes,amsmath,amssymb,twocolumn]{revtex4-2}
\usepackage{graphicx}
\usepackage{xcolor}
\usepackage{mathtools}
\usepackage{bm}
\usepackage[normalem]{ulem}
\usepackage[colorlinks=true,citecolor=blue,linkcolor=blue,urlcolor=blue,
    breaklinks=true,hypertexnames=false]{hyperref}
\IfFileExists{orcidlink.sty}{\usepackage{orcidlink}}{}
\providecommand{\orcidlink}[1]{  \href{https://orcid.org/#1}{\textnormal{\scriptsize[ORCID]}}}

\newcommand{\bea}{\begin{eqnarray}}
\newcommand{\eea}{\end{eqnarray}}

\newcommand{\vesc}{v_{\mathrm{esc}}}

\newcommand{\bfr}{{\bf r}}
\newcommand{\bfv}{{\bf v}}
\newcommand{\Ccoll}{\mathcal{C}_{\rm coll}}
\newcommand{\Ccap}{\mathcal{C}_{\rm cap}}
\newcommand{\Cann}{\mathcal{C}_{\rm ann}}
\newcommand{\Cevap}{\mathcal{C}_{\rm evap}}

\newcommand{\rv}{(\bfr,\bfv)}

\newcommand{\Kcoll}{\mathcal{K}_{\rm coll}}
\newcommand{\Kcap}{\mathcal{K}_{\rm cap}}
\newcommand{\Kann}{\mathcal{K}_{\rm ann}}
\newcommand{\Kevap}{\mathcal{K}_{\rm evap}}

\begin{document}

\title{Dark matter evolution in stellar objects: A Boltzmann transport approach}

\author{Yen-Hsun Lin
\orcidlink{0000-0001-7911-7591}}
\email{yenhsun@phys.ncku.edu.tw}
\affiliation{Institute of Physics, Academia Sinica, Taipei 115, Taiwan}

\author{Meng-Ru Wu
\orcidlink{0000-0003-4960-8706}}
\email{mwu@as.edu.tw}
\affiliation{Institute of Physics, Academia Sinica, Taipei 115, Taiwan}
\affiliation{Institute of Astronomy and Astrophysics, Academia Sinica, Taipei 106, Taiwan}
\affiliation{Physics Division, National Center for Theoretical Sciences, Taipei 106, Taiwan}

\author{Seokhoon Yun
\orcidlink{0000-0002-7960-3933}}
\email{seokhoon.yun@knu.ac.kr}
\affiliation{Department of Physics, Kyungpook National University, Daegu 41566, Korea}
\affiliation{Center for Theoretical Physics of the Universe, Institute for Basic Science (IBS), Daejeon 34126, Korea}

\begin{abstract}
Dark matter (DM) captured by stars offers a way to search for physics beyond the Standard Model. We study the evolution of captured DM with the Boltzmann transport equation (BTE), which follows both the abundance and the phase-space distribution of the bound particles from capture toward a stationary state. Averaging over orbital motion, we write the BTE as a master equation in energy and angular momentum that includes capture, scattering between bound orbits, annihilation and evaporation. Applying it to the Sun, we find that captured DM does not thermalize within the solar age when its scattering on nuclei is rare. The bound DM then remains much more extended than the isothermal distribution assumed in the classical treatment, and its annihilation rate can be suppressed by many orders of magnitude. We also apply the BTE to the inelastic Higgsino interpretation of the recent LUX-ZEPLIN result, which involves two DM components coupled through inelastic scattering. The thermal motion of solar nuclei opens capture channels that are closed for nuclei at rest and strengthens the constraint on the mass splitting by tens of keV.

\end{abstract}

\maketitle

\section{Introduction}
\label{sec:introduction}

Dark matter (DM) accounts for one-fourth of the total energy density in the Universe and plays a
central role in the formation of cosmic structure~\cite{Planck:2018vyg}.
Despite ample evidence from its gravitational effects, its particle nature and its interactions with ordinary matter remain unknown.
Capture of DM in celestial bodies connects direct searches for DM scattering~\cite{Goodman:1984dc}
to indirect searches for annihilation or decay products~\cite{Leane:2024bvh}.
Annihilation of captured DM can yield signals from the Sun, Earth and other celestial bodies~\cite{Freese:1985qw,Lin:2014hla,Silk:1985ax,Griest:1986yu,Super-Kamiokande:2015xms,ANTARES:2016xuh,IceCube:2016yoy,IceCube:2016dgk,IceCube:2025fcu,Batell:2009zp,Schuster:2009au,Bell:2011sn,Leane:2017vag,HAWC:2018szf,Mazziotta:2020foa,Bose:2021cou,Leane:2021ihh,Bhattacharjee:2023qfi,Chen:2015poa}.
DM annihilation may also power the first generation of stars and change the evolution of stars in DM-rich environments~\cite{Salati:1987ja,Fairbairn:2007bn,Scott:2008ns,Casanellas:2009dp,John:2023knt,Lopes:2021jcy,Hong:2024ozz,Ganguly:2025ffb,Spolyar:2007qv,Iocco:2008xb,Freese:2008ur,Iocco:2008rb,Taoso:2008kw,Sivertsson:2010zm,Ilie:2020nzp}.

Neutron stars (NSs) extend these probes through kinetic and annihilation heating~\cite{Kouvaris:2007ay,deLavallaz:2010wp,Kouvaris:2010vv,Baryakhtar:2017dbj,Raj:2017wrv,Bell:2018pkk,Chen:2018ohx,Acevedo:2019agu,Lin:2021hen} that alters their cooling curves. Asymmetric DM can accumulate without annihilating and trigger gravitational collapse that engulfs the entire NS~\cite{Goldman:1989nd,Kouvaris:2011fi,McDermott:2011jp,Bell:2013xk,Garani:2018kkd,Lin:2020zmm}.
White dwarfs (WDs) in environments that may be rich in DM, such as globular clusters and the Galactic center, provide another probe, since capture and annihilation can heat them and their observed luminosities constrain the scattering on nuclei~\cite{Moskalenko:2007ak,Bertone:2007ae,Hurst:2014uda,Bell:2021fye}.
These studies have been extended to inelastic DM~\cite{McCullough:2010ai} and to the capture and thermalization of heavy particles~\cite{Bell:2024qmj}, and annihilation into long-lived mediators that escape WDs can also produce gamma-ray signals~\cite{Acevedo:2023xnu}.
All these signals depend on how much DM is bound and how it is distributed inside the host, so their interpretation requires following the captured population as it evolves.

The classical capture formalism of Refs.~\cite{Press:1985ug,Gould:1987ir} describes
halo DM scattering into bound orbits, with extensions to multiple scattering~\cite{Gould:1991va,Bramante:2017xlb,Dasgupta:2019juq,Ilie:2021iyh},
more general interactions~\cite{Liang:2013dsa,Catena:2015uha,Dasgupta:2020dik},
and different objects and kinematic regimes~\cite{Bringmann:2018lay,Kozar:2021iur,Leane:2023woh}.
The classical treatment evolves the particle number using evaporation and
annihilation coefficients evaluated for an isothermal distribution~\cite{Griest:1986yu,Busoni:2013kaa,Garani:2017jcj,Busoni:2017mhe,Garani:2021feo}, assuming that the captured DM thermalizes quickly with the stellar nuclei.
Weak scattering~\cite{Peter:2009mk,Widmark:2017yvd} or inelastic thresholds~\cite{Nussinov:2009ft,Menon:2009qj,Blennow:2018xwu}
can, however, delay thermalization beyond the stellar age.
Since evaporation is sensitive to the weakly bound population and annihilation
depends on the spatial overlap of particles, the abundance and the distribution must then be determined together.

\begin{figure*}
    \centering
    \includegraphics[width=.96\textwidth]{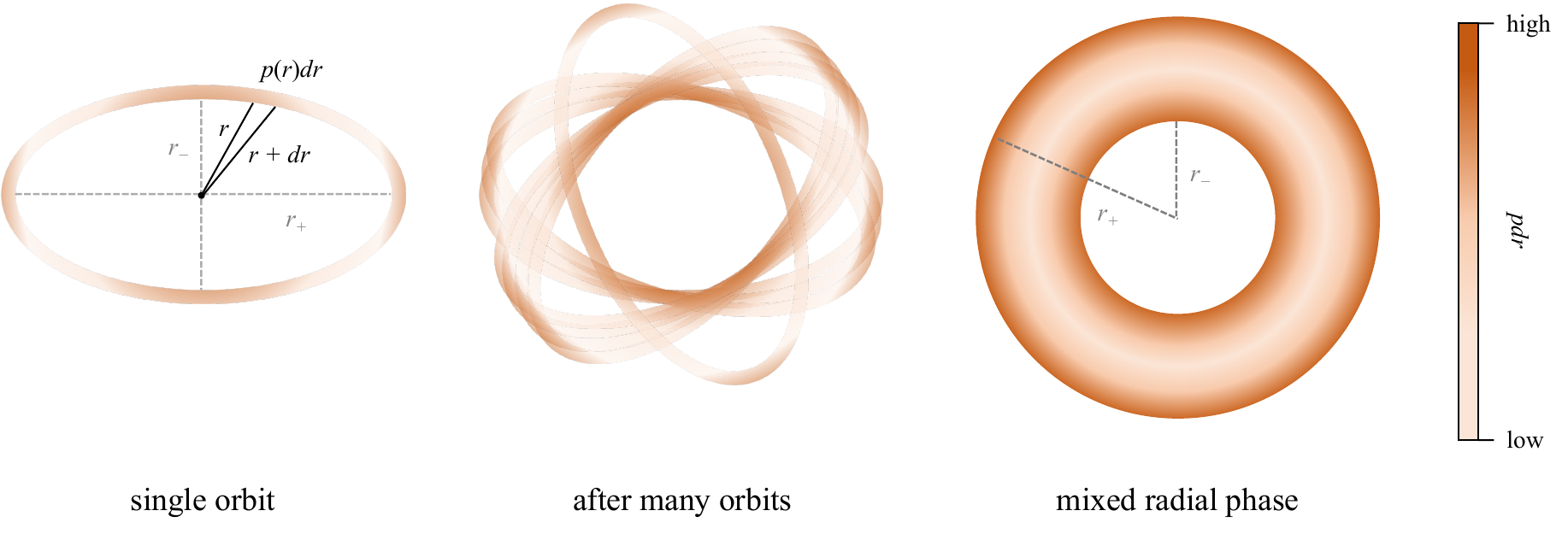}
    \caption{Left: Radial probability along a bound orbit. Middle: In the solar potential,
a bound orbit generally precesses and does not close after one radial period
$\tau_r$. Right: An ensemble with mixed radial phases fills an annulus in the
orbital plane, with radial probability proportional to the time spent at each radius.
}
    \label{fig:radial_prob}
\end{figure*}

Earlier studies have used the Boltzmann transport equation (BTE) and numerical phase-space evolution to
investigate selected processes separately, including the redistribution of bound DM among orbits by scattering on nuclei, evaporation and heat
transport~\cite{Gould:1987ju,Spergel:1984re,Gould:1989hm,Gould:1989ez,Vincent:2013lua,Vincent:2015gqa,Vincent:2016dcp,Nauenberg:1986em,Liang:2016yjf,Liang:2018cjn,Banks:2021sba,Banks:2024eag,Blennow:2018xwu}.
In this work, we evolve capture, scattering between bound orbits, annihilation and evaporation together.
When a bound particle completes many orbits between collisions, the captured population can be described by the number of particles on each orbit, and collisions move particles from one orbit to another.
Following this population determines both the abundance and the distribution of captured DM without assuming thermalization, and integrating over orbits recovers the classical number evolution equation with coefficients set by the evolving distribution.

Throughout the paper, we take the Sun, described by the AGSS09 solar model~\cite{Asplund:2009fu,Serenelli:2009yc}, as an example and consider spin-independent (SI) and spin-dependent (SD) DM-nucleon scattering, although the same approach applies to other stellar objects.
Since each process enters as a separate term, DM with several components can be treated by adding the terms that couple them, such as scattering between the components or the conversion of one into another~\cite{Chen:2018lsk}. We demonstrate this multicomponent case with the inelastic Higgsino model, which has been proposed as one possible interpretation~\cite{Wu:2026nhi,DiMauro:2026ldr} of the recent LUX-ZEPLIN (LZ) result~\cite{LZ:2026axp}.

The paper is organized as follows. In Sec.~\ref{sec:general_formalism}, we introduce the BTE in position and velocity space, and in Sec.~\ref{sec:orbital-space} we describe the orbital phase space, state the orbit-averaging approximation and reconstruct the spatial density. Sec.~\ref{sec:master} derives the master equation and its connection to the classical number equation, Secs.~\ref{sec:numerical_results} and~\ref{sec:multicomponent} apply the BTE to DM in the Sun and to the inelastic Higgsino model, respectively, and Sec.~\ref{sec:summary} summarizes.
Natural units $c=\hbar=k_B=1$ are used unless specified otherwise.

\section{General formalism}
\label{sec:general_formalism}

The captured DM population is governed by the BTE,
\begin{equation}
  \frac{\partial f}{\partial t}
  +{\bf v}\cdot\nabla_{{\bf r}}f
  -\nabla_{{\bf r}}\Phi\cdot\nabla_{{\bf v}}f
  =\mathcal C[f],
  \label{eq:boltzmann}
\end{equation}
where $f=f({\bf r},{\bf v},t)$ is the phase-space distribution of gravitationally
bound DM, defined such that $f\,d^3{\bf r}d^3{\bf v}$ is the number of particles in the phase-space volume $d^3{\bf r}d^3{\bf v}$, and the two advection terms describe free motion and gravitational
acceleration in the gravitational potential per unit mass, $\Phi$.
We assume a static, spherically symmetric Sun whose nuclei have isotropic thermal velocities.
The interaction operator
$\mathcal C=\Ccap+\Ccoll+\Cevap+\Cann$ contains three contributions from scattering on nuclei and one from annihilation.
Scattering on nuclei can capture halo DM onto bound orbits, redistribute bound DM among orbits in phase space, and make bound DM evaporate by giving it enough energy to escape.
The spatial number density of bound DM is
\begin{equation}
  n_\chi({\bf r},t)=\int d^3{\bf v}\,f({\bf r},{\bf v},t),
\end{equation}
which yields the total population
\begin{equation}
  N_\chi(t)=\int d^3{\bf r}d^3{\bf v}\,f({\bf r},{\bf v},t).
  \label{eq:total_Nx}
\end{equation}
Bound orbits can extend beyond the solar surface, so $n_\chi(\bfr,t)$ need not
vanish outside the Sun. We suppress explicit time arguments when they are not needed.

\subsection{DM-nucleus scattering operators}\label{sec:Ccoll}\label{sec:continuous-capture}\label{sec:evaporation}

Scattering on a nucleus changes the velocity of a DM particle at fixed position.
A particle with velocity $\bfv$ at radius $r$ scatters to velocity $\bfv'$ with the rate per unit $d^3\bfv'$
\begin{widetext}
\begin{equation}
  R({\bf v}\to{\bf v}',r)
  =\sum_A n_A(r)\int d\Omega_u\int_0^\infty du\,u^2f_A(r,u)
  \frac{\sigma_{\chi A}}{4\pi}\int d\Omega_{g'}g F_A^2(q)
  \delta^{(3)}({\bf v}'-{\bf v}'_f),
  \label{eq:R_v2v'}
\end{equation}
\end{widetext}
where $n_A(r)$ is the number density of nuclear species $A$, $f_A(r,u)$ is the thermal velocity distribution of the nuclei, with ${\bf u}$ the velocity of a nucleus, $g=|\bfv-{\bf u}|$ is the relative speed and $\bfv_f'$ is the outgoing DM velocity.
We take the DM-nucleus cross section $\sigma_{\chi A}$ to be velocity independent,
and $F_A(q)$ is the nuclear form factor at momentum transfer $q=m_\chi|{\bf v}_f'-{\bf v}|$. Because the Sun is spherically symmetric, the rate depends on position only through $r$.
Elastic scattering preserves the relative speed, $g'=g$, and
the incident velocities together with the outgoing relative direction $\Omega_{g'}$ fix
$\bfv'_f$ through energy and momentum conservation, as given in App.~\ref{app:kinematics}.

The role of a collision depends on whether the initial and final speeds lie below the escape speed $\vesc(r)$ at $r$.
Scattering between bound states, $v<\vesc(r)$ and $v'<\vesc(r)$, redistributes DM in phase space,
\begin{widetext}
\begin{equation}
  \Ccoll[f]
  =\int_{v'<\vesc(r)}d^3\bfv'\,[R({\bf v}'\to{\bf v},r)f({\bf r},{\bf v}')
  -R({\bf v}\to{\bf v}',r)f({\bf r},{\bf v})],
  \label{eq:Ccoll_f}
\end{equation}
\end{widetext}
where the two terms describe particles scattered into and out of the state with velocity $\bfv$.
Scattering of a bound particle to $v'>\vesc(r)$ gives evaporation,
\begin{equation}
  \Cevap[f]
  =-f\rv \int_{v'>\vesc(r)}d^3\bfv'\,R({\bf v}\to{\bf v}',r),
  \label{eq:Cevap}
\end{equation}
where we assume that an evaporated particle leaves the Sun without another collision with a nucleus.

Scattering of halo DM into a bound state gives capture.
A halo particle with speed $v_\infty$ far from the Sun reaches radius $r$ with speed $w(r)=\sqrt{v_\infty^2+\vesc^2(r)}$
and is captured if its outgoing speed satisfies $v<\vesc(r)$.
The capture operator is then
\begin{align}
  \Ccap[f]
  &=n_\infty\int_0^\infty dv_\infty\,
  \frac{f_\odot(v_\infty)}{v_\infty}w(r)\nonumber\\
  &\qquad\times\int\frac{d\Omega_w}{4\pi}\,R({\bf w}\to{\bf v},r),
  \label{eq:local-capture}
\end{align}
where $n_\infty$ is the halo DM number density far from the Sun,
$f_\odot(v_\infty)$ is the halo speed distribution in the solar frame,
and $d\Omega_w/4\pi$ averages over the incident directions.
Since we neglect DM self-capture~\cite{Zentner:2009is} in this work, capture is an external source that does not depend on the bound
population.

\subsection{Annihilation}
\label{sec:annihilation}

For $s$-wave annihilation, $\sigma_{\rm ann}v_{\rm rel}$ does not depend on the relative speed $v_{\rm rel}$ of the annihilating pair, and we denote this
constant by $\langle\sigma v\rangle$.
The
annihilation operator then depends only on the local density of bound partners,
\begin{equation}
  \Cann[f]=-\langle \sigma v\rangle n_\chi({\bf r})f\rv.
  \label{eq:local-annihilation}
\end{equation}
A velocity-dependent annihilation cross section
would instead require an average over the
velocities of both particles.

\section{BTE in orbital phase space}\label{sec:orbital-space}

Between collisions, a bound particle moves on a fixed orbit in the static potential, and its specific energy $E$ and angular-momentum
magnitude $L$ are conserved.
When the mean time between collisions is much longer than the radial period, $\tau_{\rm coll}\gg\tau_r$, a particle completes many orbits before it scatters again.
We then assume that bound DM is distributed uniformly in radial phase, the time since the last pericenter passage in units of the radial period, along each orbit and describe the captured population by the number of particles per orbit $\xi=(E,L)$.
Below, we define the orbital coordinates and the orbital number density, and reconstruct the radial density with this approximation.

\subsection{Orbital coordinates}

With the gravitational potential set to zero at infinity, $\Phi(\infty)=0$, the specific energy and angular momentum of a phase-space point $\rv$ are
\begin{equation}
  E=\frac{v^2}{2}+\Phi(r),\qquad L=|{\bf r}\times{\bf v}|,
  \label{eq:E_and_L}
\end{equation}
and the bound orbits that cross the solar interior satisfy
\begin{equation}
  \Phi(0)<E<0,\qquad 0<L<L_\odot(E),
  \label{eq:bound_and_cross_condition}
\end{equation}
where
\begin{equation}
  L_\odot(E)=\max_{\substack{0\leq r\leq R_\odot\\\Phi(r)\leq E}}
  r\sqrt{2[E-\Phi(r)]}.
\end{equation}
This domain includes orbits confined to the Sun and those extending beyond its surface, and
the boundary is discussed in App.~\ref{app:solar_cross_orbit}.
We write $\xi\rv\equiv(E,L)$ for the orbital coordinates of a phase-space point given by Eq.~\eqref{eq:E_and_L}.
At radius $r$, the radial and tangential speeds, $v_r$ and $v_t$ respectively, are
\begin{equation}
  v_r=\sqrt{2[E-\Phi(r)]-\frac{L^2}{r^2}},\qquad v_t=\frac{L}{r},
  \label{eq:vr_vt}
\end{equation}
with total speed $v=\sqrt{v_r^2+v_t^2}$.
Throughout this work, we take $v_r\geq0$ and count the inward and outward passages together, since their
radial speeds are equal in a spherically symmetric potential.
A bound particle moves from the pericenter $r_-$ to the apocenter $r_+$ and back within the radial period
\begin{equation}
\tau_r(\xi)= 2\int_{r_-}^{r_+}\frac{dr}{v_r(r,\xi)}.
\end{equation}

\subsection{Orbital number density}\label{sec:orbit_average}

The number density of bound particles per $dEdL$, derived together with the other quantities of this subsection in App.~\ref{app:rv_to_EL}, is
\begin{equation}
  n_\chi(\xi)=\int d^3\bfr d^3\bfv\,\Delta(\xi-\xi\rv)\,f\rv,
  \label{eq:nx_def}
\end{equation}
where $\Delta(\xi-\xi\rv)\equiv\delta(E-E\rv)\delta(L-L\rv)$
selects the phase-space points $\rv$ whose orbital coordinates $\xi\rv$ equal $\xi$. Since $E$ and $L$ are defined per unit mass, $\Delta$ has the dimension of $1/(rv^3)$, and $n_\chi(\xi)$ is the number of particles per $dEdL$.
The phase-space volume of the orbits in $dEdL$ is $dV_\xi = 8\pi^2 L \tau_r(\xi) dEdL$, and Eq.~\eqref{eq:nx_def} becomes
\begin{equation}
    n_\chi(\xi) = 8\pi^2 L \tau_r(\xi) f(\xi),\label{eq:nx_xi}
\end{equation}
where $f(\xi)$ is the phase-space distribution averaged over the orientations and radial phases of the orbit.
The total bound population is
\begin{equation}
  N_\chi = \int n_\chi(\xi) dEdL= \int f(\xi) dV_\xi.
  \label{eq:dNx_xi}
\end{equation}
Because the interaction rates vary along the orbit with the solar profile and the DM density, we evaluate them with the orbit-averaging approximation,
\begin{equation}
  f(\bfr,\bfv,t)\simeq f(\xi\rv,t),
  \label{eq:orbit_average}
\end{equation}
which requires $\tau_{\rm coll}\gg\tau_r$ and a uniform spread of the particles in radial phase.
The latter holds because particles reach an orbit at random times through capture or scattering.

\begin{figure}
\centering
    \includegraphics[width=.85\columnwidth]{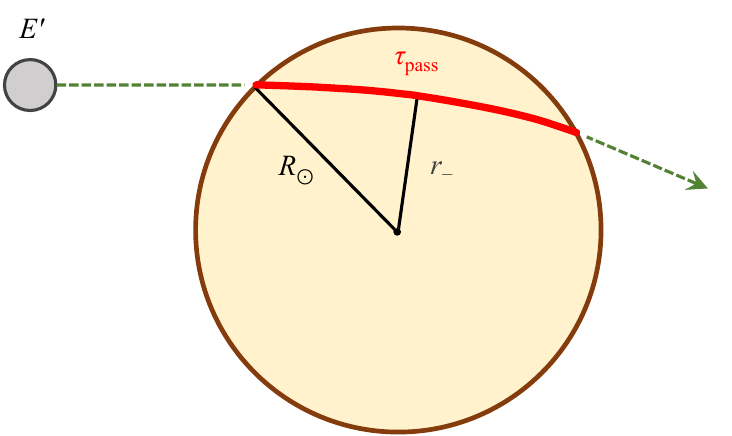}
    \caption{Passage of an unbound DM particle through the solar interior (thick red line), with incident specific energy $E'>0$.}
    \label{fig:tau_pass}
\end{figure}

\begin{figure}[t]
\centering
 \includegraphics[width=.48\textwidth]{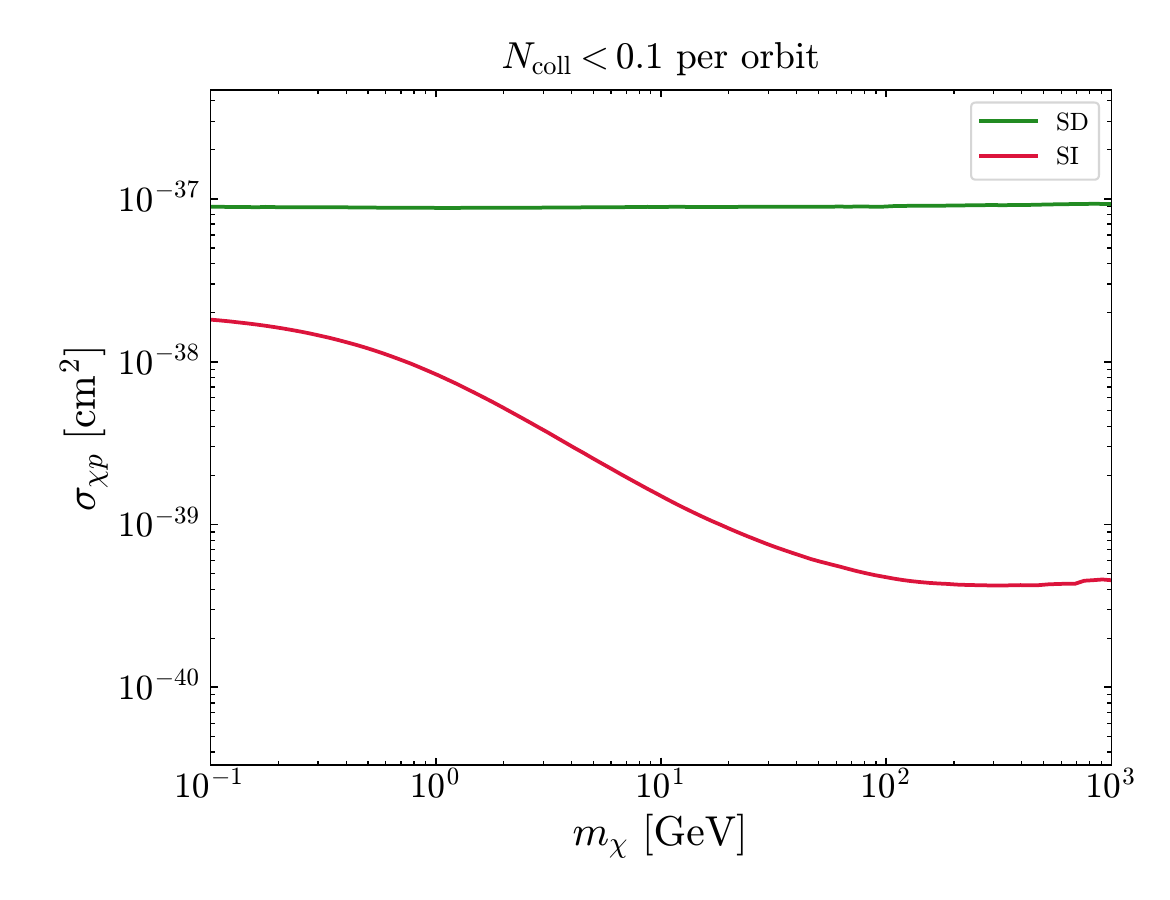}
 \caption{
 Below the SD (green) and SI (red) curves, 99\% of the bound DM on our $(E,L)$ grid has $N_{\rm coll}<0.1$ per orbit at all times of the evolution.}
 \label{fig:bte_validity}
\end{figure}

\subsection{Reconstruction of radial DM number density}\label{sec:density}

\begin{figure*}
 \centering
 \includegraphics[width=.48\textwidth]{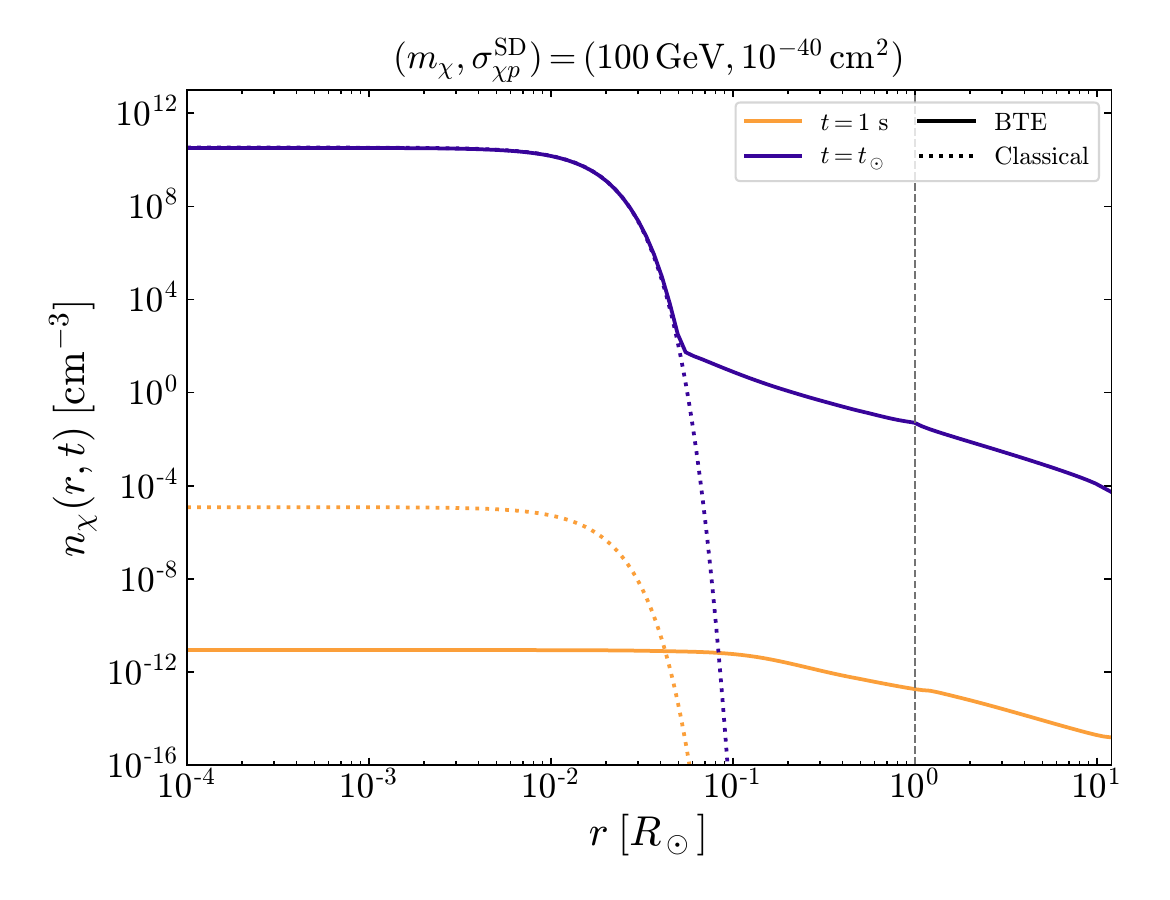}\quad
 \includegraphics[width=.48\textwidth]{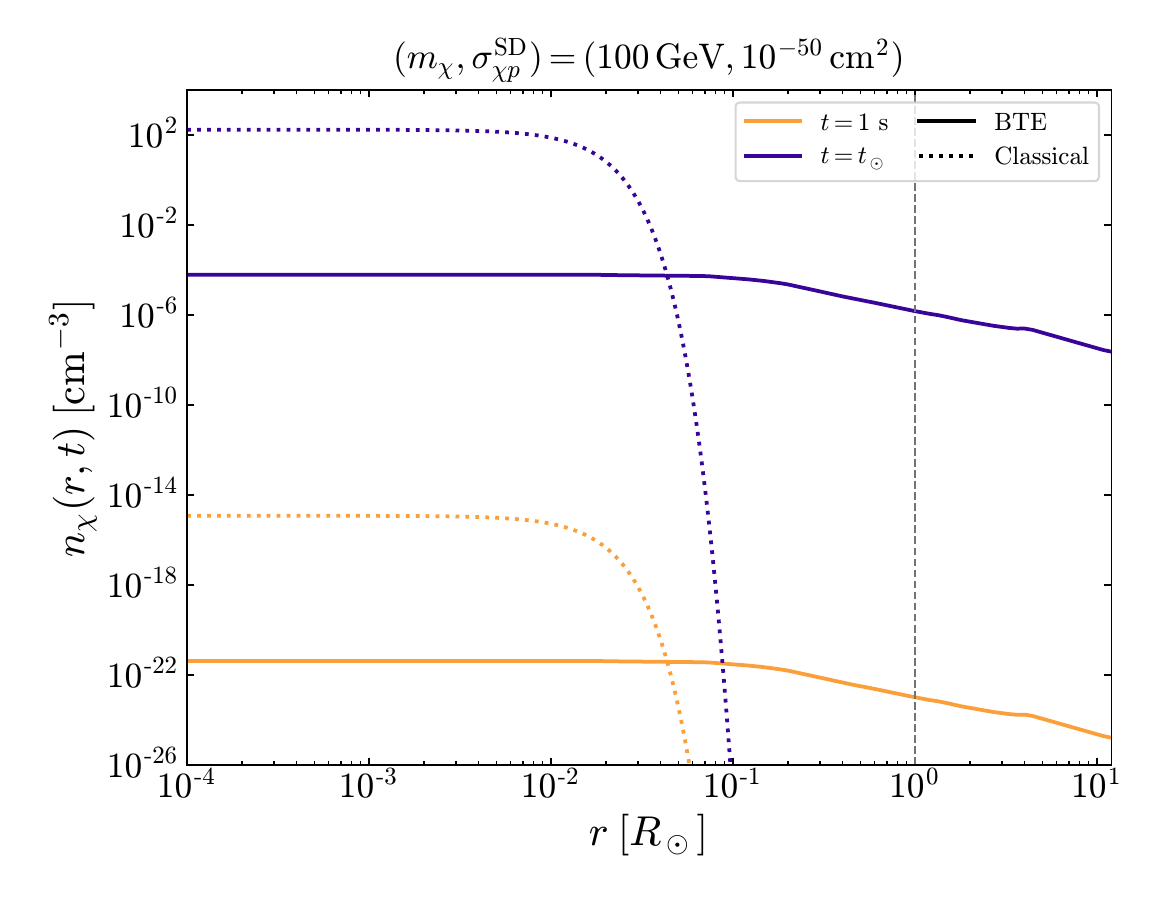}
 \caption{$n_\chi(r,t)$
 for the SD benchmark
 $m_\chi=100~{\rm GeV}$, with
 $\sigma_{\chi p}=10^{-40}~{\rm cm^2}$ (left) and
 $10^{-50}~{\rm cm^2}$ (right).
 Orange curves show the distribution of DM captured within one second, $n_\chi(\xi)=\Kcap(\xi)\times1$~s, and dark-blue curves the
  distribution at $t=t_\odot$. Solid curves are the BTE results
  and dotted curves the classical isothermal results under continuous capture
  and annihilation. Vertical dashed lines mark the solar surface.}

\label{fig:nx_evolution}
\end{figure*}

A particle on orbit $\xi$ spends a time $dr/v_r$ in a radial interval on each passage, so the probability of finding it between $r$ and $r+dr$ is
\begin{equation}
  p(r,\xi)dr=\frac{2}{\tau_r(\xi)}\frac{dr}{v_r(r,\xi)},\label{eq:pr_xi}
\end{equation}
with $p=0$ outside $r_-<r<r_+$.
This probability is enhanced near the turning points $r_\pm$, where $v_r\to0$.
To obtain a spherically symmetric density, we further assume that the orbits are isotropically oriented.
Precession randomizes the pericenter direction within each orbital plane, as shown in the middle panel of Fig.~\ref{fig:radial_prob}.
The orientation of the orbital plane changes only through collisions, and repeated collisions erase any anisotropy inherited from the halo flux. Even when collisions are too rare to erase this anisotropy, the assumption of isotropically oriented orbits remains a good approximation, as we show in App.~\ref{app:anisotropy}.
With these assumptions, the radial number density is
\begin{equation}
  n_\chi(r)=\frac{1}{4\pi r^2}
  \int_{\Phi(0)}^0dE\int_0^{L_\odot(E)}dL\,n_\chi(\xi)p(r,\xi),\label{eq:nx(r)_reconstruction}
\end{equation}
as illustrated in Fig.~\ref{fig:radial_prob}.
Integrating over all radii recovers the total bound population,
\begin{equation}
  N_\chi=\int_0^\infty dr\, 4\pi r^2 n_\chi(r),
  \label{eq:Nx_r_int}
\end{equation}
which is equivalent to Eq.~\eqref{eq:total_Nx}.

\section{Master equation for DM evolution in orbital phase space}\label{sec:master}

Multiplying Eq.~\eqref{eq:boltzmann} by $\Delta(\xi-\xi\rv)$ and integrating over phase space, we obtain the master equation for the evolution of $n_\chi(\xi)$,
\begin{align}\label{eq:master_eq}
  \frac{\partial n_\chi(\xi)}{\partial t}&=
 \int d^3\bfr d^3\bfv\,\Delta(\xi-\xi\rv)\, \mathcal{C}[f]\rv \nonumber \\
 &
 \equiv \mathcal{K}[n_\chi](\xi).
\end{align}
The advection terms drop out without approximation in this projection, because the motion between collisions conserves $E$ and $L$ and preserves phase-space volume (App.~\ref{app:0_advection}).
The
kernel $\mathcal K$ gives the rate of population change per
$dEdL$, and we derive the master equation and the collisional kernel discussed below in Apps.~\ref{app:master_derivation} and \ref{app:collisional_kernel}, respectively.

\subsection{DM-nucleus scattering kernels}\label{sec:Kcoll}\label{sec:Kcap}

\begin{figure*}[t]
    \centering
    \includegraphics[width=.48\textwidth]{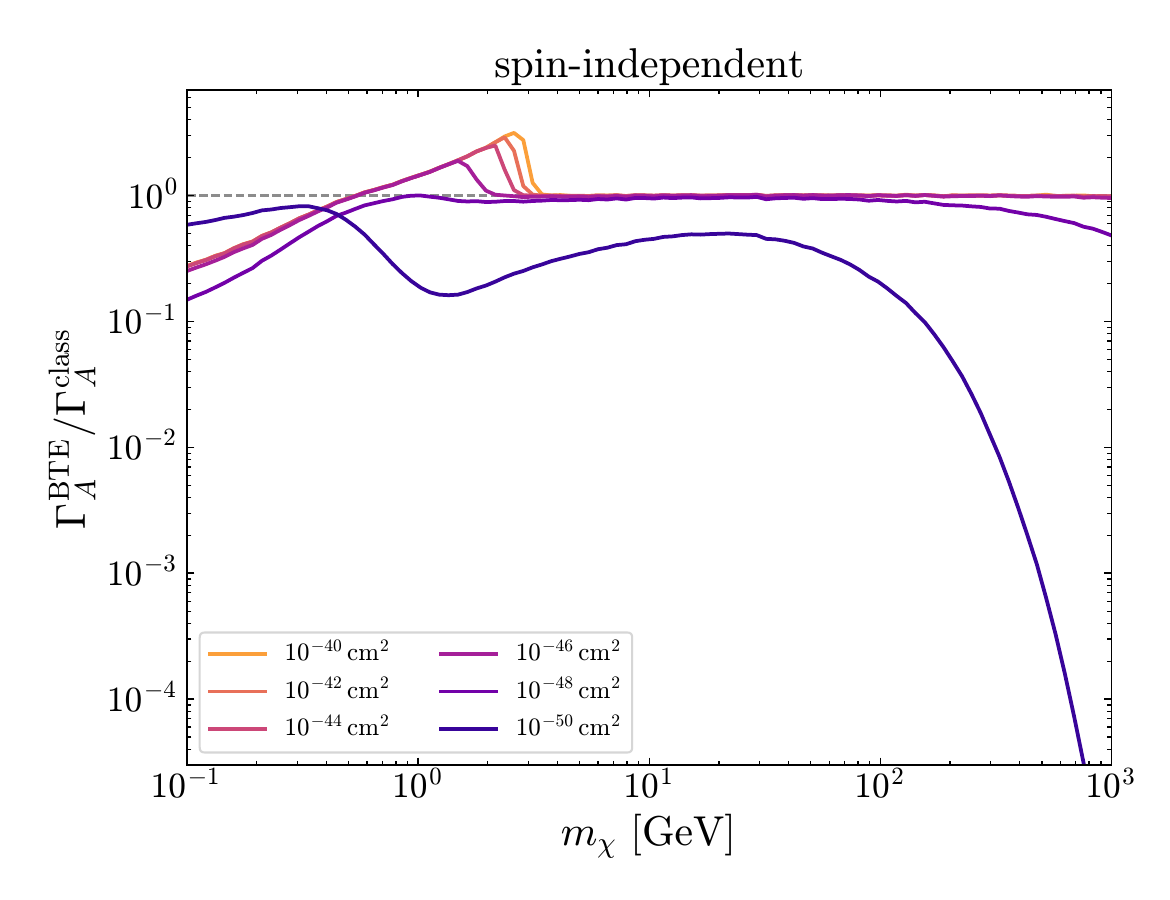}~
    \includegraphics[width=.48\textwidth]{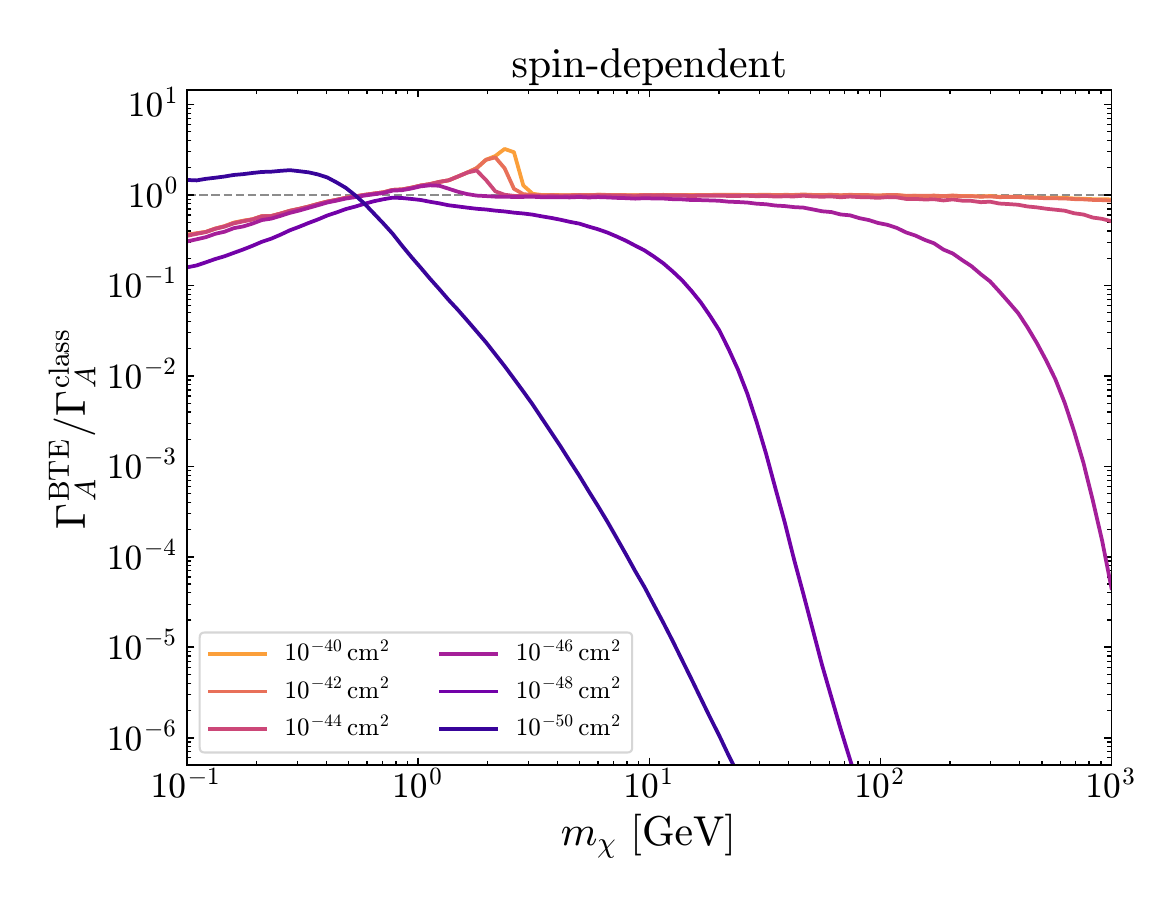}
    \caption{BTE-to-classical ratio of the annihilation rate inside the Sun at $t_\odot$,
    for SI (left) and SD (right) interactions.
    Colors indicate the different cross sections.
    }
    \label{fig:Gamma_A_ratio}
\end{figure*}

The rate per particle to scatter from orbit $\xi$ into the orbits within $dE'dL'$ around $\xi'=(E',L')$ is $K(\xi\to\xi')\,dE'dL'$.
Averaging the local rate $R$, Eq.~\eqref{eq:R_v2v'}, along the incident orbit
and selecting the final energy $E_f'$ and angular momentum $L_f'$, given by Eq.~\eqref{eq:outgoing-orbit}, gives
\begin{widetext}
\begin{equation}
  K(\xi\to\xi')
  =\frac{2}{\tau_r(\xi)}\int_{r_-}^{r_+}\frac{dr}{v_r(r,\xi)}
 \sum_A n_A(r)\frac{\sigma_{\chi A}}{4\pi}
  \int_0^\infty du\,u^2f_A(r,u)\int d\Omega_u
  \int d\Omega_{g'}gF_A^2(q)
  \delta(E'-E_f')\delta(L'-L_f').
  \label{eq:rate_K}
\end{equation}
\end{widetext}
Since \mbox{$n_A(r>R_\odot)=0$}, only the solar interior contributes to the radial integral.
The average is nevertheless normalized by the full period $\tau_r(\xi)$ to account
for the time a particle may spend outside the Sun between collisions.
As in Sec.~\ref{sec:Ccoll}, the role of a collision depends on whether the initial and final orbits are bound.
Scattering between bound orbits redistributes the population,
\begin{widetext}
\begin{align}
  \Kcoll[n_\chi](\xi)&=
  \int d^3\bfr d^3\bfv\,\Delta(\xi-\xi\rv)\, \Ccoll[f]\rv \nonumber \\
  &=\int_{\Phi(0)}^0dE'\int_0^{L_\odot(E')}dL'[K(\xi'\to\xi)n_\chi(\xi')
  -K(\xi\to\xi')n_\chi(\xi)].
  \label{eq:Kcoll}
\end{align}
\end{widetext}
Otherwise, the particle becomes unbound, which gives the evaporation kernel,
\begin{gather}
  \Kevap[n_\chi](\xi)
  =-n_\chi(\xi) \qquad\qquad\qquad \nonumber \\
  \qquad\qquad\qquad\times \int_0^\infty dE'\int_0^{L_\odot(E')}dL'K(\xi\to\xi'),
  \label{eq:orbital-evaporation}
\end{gather}
where the integration over $E'>0$ selects the collisions that eject the particle.

\begin{figure*}[t]
    \centering
    \includegraphics[width=.48\textwidth]{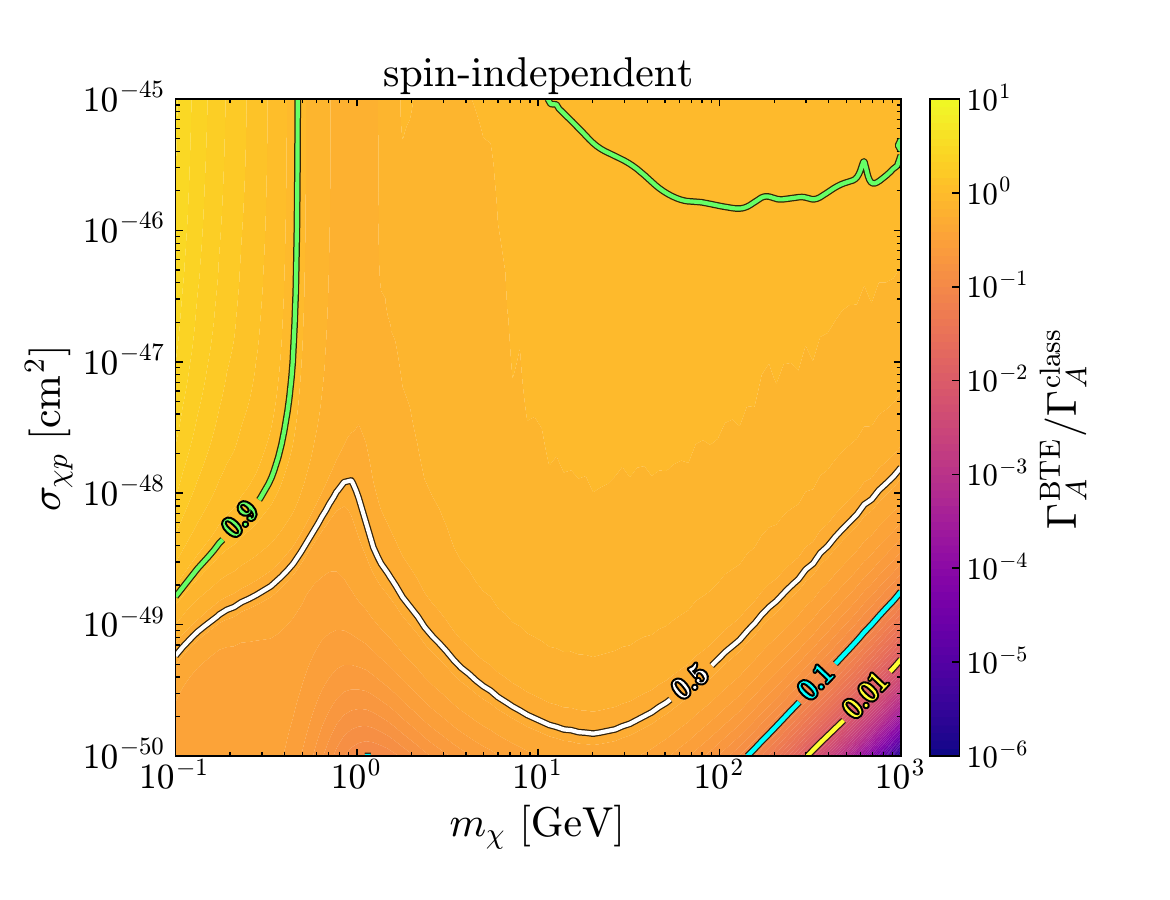}~
    \includegraphics[width=.48\textwidth]{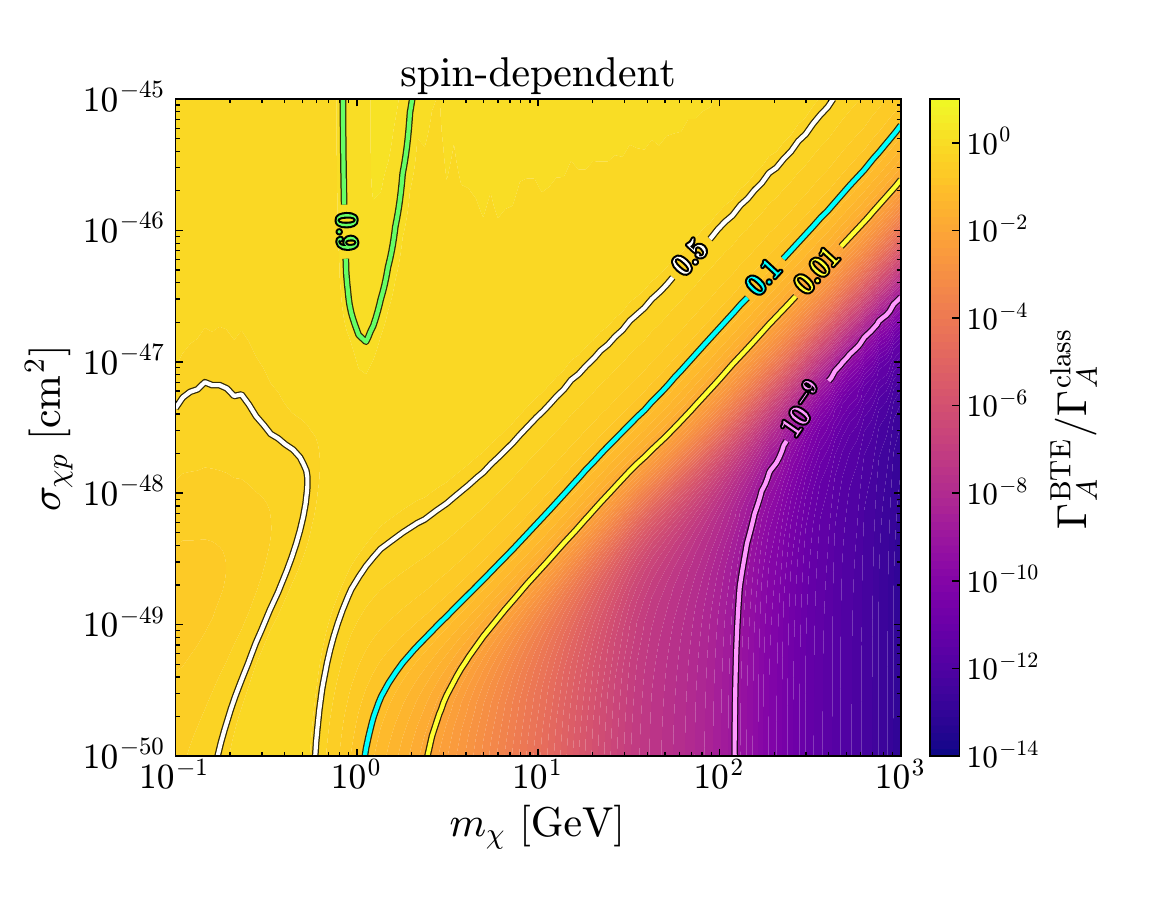}
    \caption{Similar to Fig.~\ref{fig:Gamma_A_ratio}, with a scan of $\Gamma_A^{\rm BTE}/\Gamma_A^{\rm class}$ over the $m_\chi$-$\sigma_{\chi p}$ plane
    for SI (left) and SD (right) interactions. The contours indicate ratios of 0.9 (green), 0.5 (white), 0.1 (cyan), 0.01 (yellow), and $10^{-9}$ (pink). 
    }
    \label{fig:Gamma_A_ratio_scan}
\end{figure*}

Capture moves particles from unbound to bound orbits.
An unbound particle with incident energy $E'=v_\infty^2/2>0$ at infinity crosses the solar interior once, with the passage time
\begin{equation}
\tau_{\rm pass}(\xi')=2\int_{r_-(\xi')}^{R_\odot}
  \frac{dr}{v_r(r,\xi')},
  \label{eq:halo-transit-time}
\end{equation}
as illustrated in Fig.~\ref{fig:tau_pass}.
Averaging the scattering rate over this passage gives the source of bound DM,
\begin{align}
  \Kcap[n_\chi](\xi)
  &=2\pi n_\infty\int_0^\infty dE'
  \int_0^{L_\odot(E')}dL'\nonumber\\
  &\quad\times L'\tau_{\rm pass}(\xi')
  \frac{f_\odot(\sqrt{2E'})}{2E'}K(\xi'\to\xi),
  \label{eq:orbital-capture-source}
\end{align}
where $K(\xi'\to\xi)$ is given by Eq.~\eqref{eq:rate_K} with $\tau_{\rm pass}$ in place of $\tau_r$
and with the radial integral ending at $R_\odot$. The integral over $E'$ includes all
incident unbound energies.

\begin{figure*}[t]
    \centering
    \includegraphics[width=.48\textwidth]{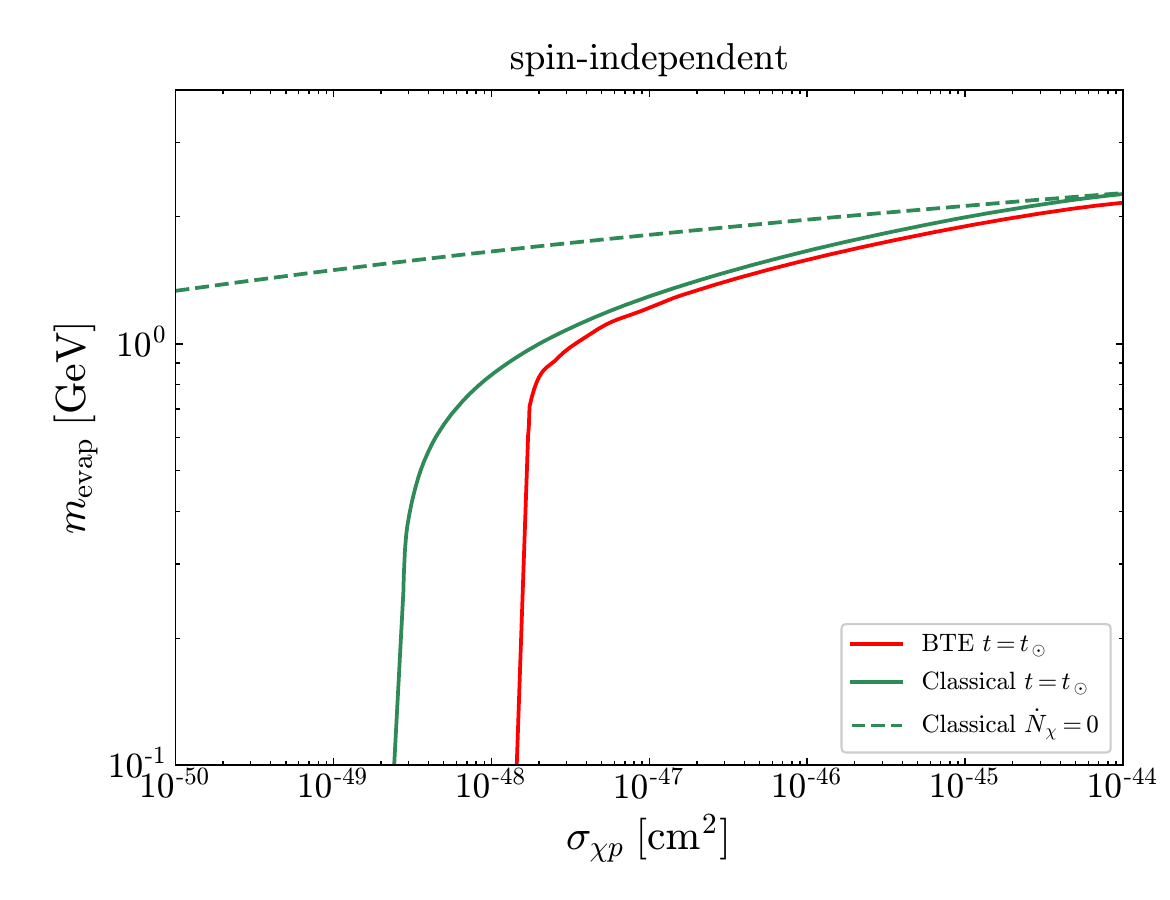}~
    \includegraphics[width=.48\textwidth]{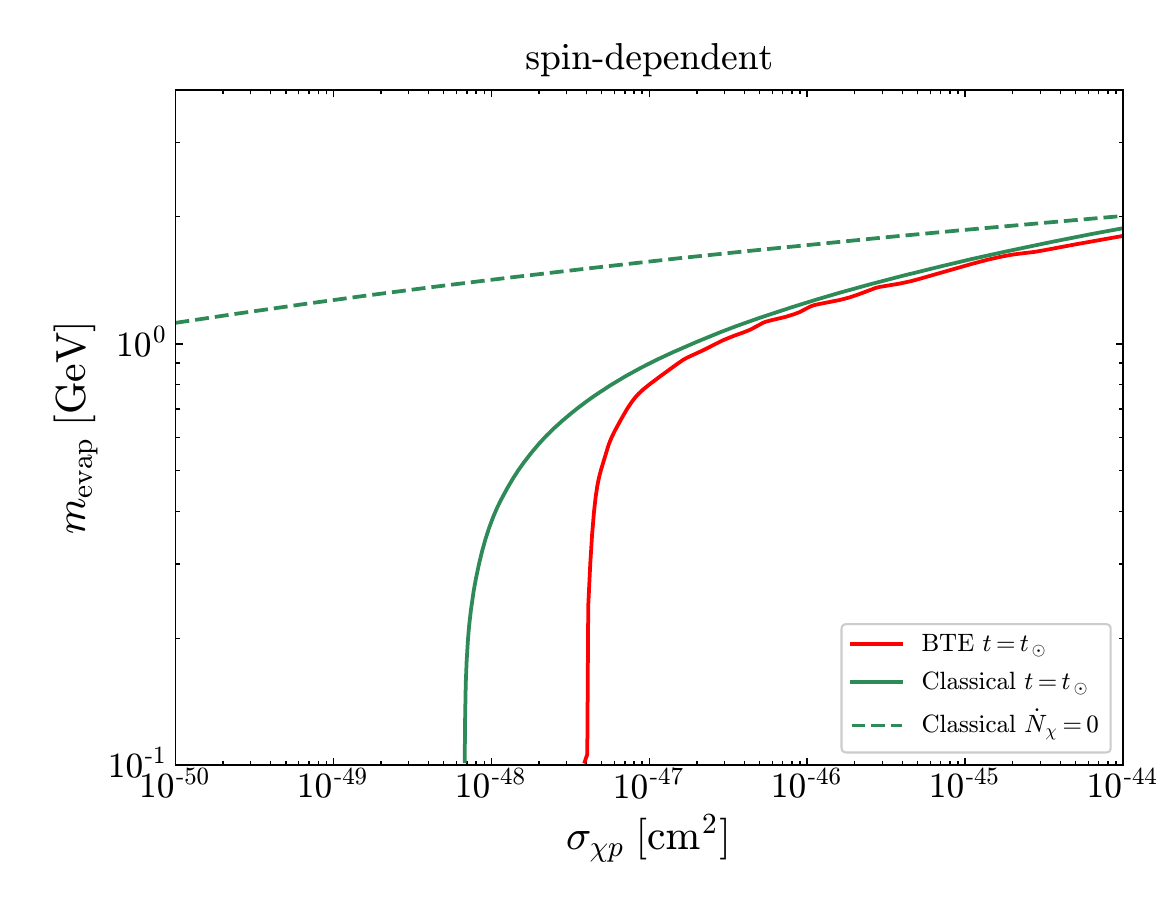}
    \caption{Evaporation mass defined by Eq.~\eqref{eq:evaporation_mass}
    for SI (left) and SD (right) scattering.
    Red and green solid curves show the BTE and classical results at $t_\odot$, and
    green dashed curves show the classical stationary solution with $\dot N_\chi=0$.
    }

\label{fig:m_evap}
\end{figure*}

\subsection{Annihilation kernel}
\label{sec:Kann-Kevap}

Annihilation couples the populations of different bound orbits through their
spatial overlap.
Averaging the local loss rate $\langle\sigma v\rangle n_\chi(r)$ over the orbit with the radial probability of Eq.~\eqref{eq:pr_xi} and using Eq.~\eqref{eq:nx(r)_reconstruction},
we obtain
\begin{align}
  \Kann[n_\chi](\xi)
  &=-n_\chi(\xi)
\int_{\Phi(0)}^0dE'\int_0^{L_\odot(E')}dL'n_\chi(\xi')\nonumber\\
  &\quad\times \langle \sigma v\rangle\int_0^\infty dr\,\frac{p(r,\xi)p(r,\xi')}{4\pi r^2},
  \label{eq:Kann}
\end{align}
where the factor $1/(4\pi r^2)$ distributes the radial probability over a spherical shell.

\subsection{Connection to the classical treatment of DM evolution}

Integrating Eq.~\eqref{eq:master_eq} over the bound orbital phase space gives the number evolution equation
\begin{equation}
\frac{dN_\chi}{dt}= C_c + C_e N_\chi + C_a N_\chi^2,
\label{eq:evol_classical}
\end{equation}
where the capture, annihilation and evaporation coefficients are given by the orbital kernels,
\begin{subequations}
\begin{align}
C_c &= \int dEdL\,\Kcap[n_\chi](\xi),\label{eq:Cc} \\
C_a &= \frac{1}{N_\chi^2}\int dEdL\,\Kann[n_\chi](\xi),\label{eq:Ca} \\
C_e &= \frac{1}{N_\chi}\int dEdL\,\Kevap[n_\chi](\xi). \label{eq:Ce}
\end{align}
\end{subequations}
The integrals cover $\Phi(0)\leq E\leq0$ and $0\leq L\leq L_\odot(E)$, and
with this convention $C_a$ and $C_e$ are negative. Although capture does not depend on the bound population $n_\chi$, we write the capture kernels with an argument, such as $\Kcap[n_\chi]$, to keep the notation uniform.
Bound-to-bound scattering only redistributes particles in phase space and does not change $N_\chi$, so it vanishes in the number evolution equation.
The classical treatment assumes that the distribution rapidly approaches a
stationary, isothermal form, $n_\chi^{\rm iso}(r)\propto\exp(-m_\chi\Phi(r)/T_\chi)$, where the DM temperature $T_\chi$ is computed following Ref.~\cite{Garani:2017jcj}, making these coefficients time independent. For this profile, about half of the DM lies within the thermal radius $r_{\rm th}=\sqrt{3T_\odot(0)/(2\pi G\rho_c m_\chi)}$, where $G$ is Newton's constant and $T_\odot(0)$ and $\rho_c$ are the temperature and density at the solar center.
Frequent collisions with nuclei can drive DM toward such a distribution in the
solar core. At small scattering cross sections, however, relaxation can take
longer than the solar age, and the evolving distribution must be obtained from
the master equation.

\section{Numerical results}
\label{sec:numerical_results}

\begin{figure*}
    \centering
    \includegraphics[width=.48\textwidth]{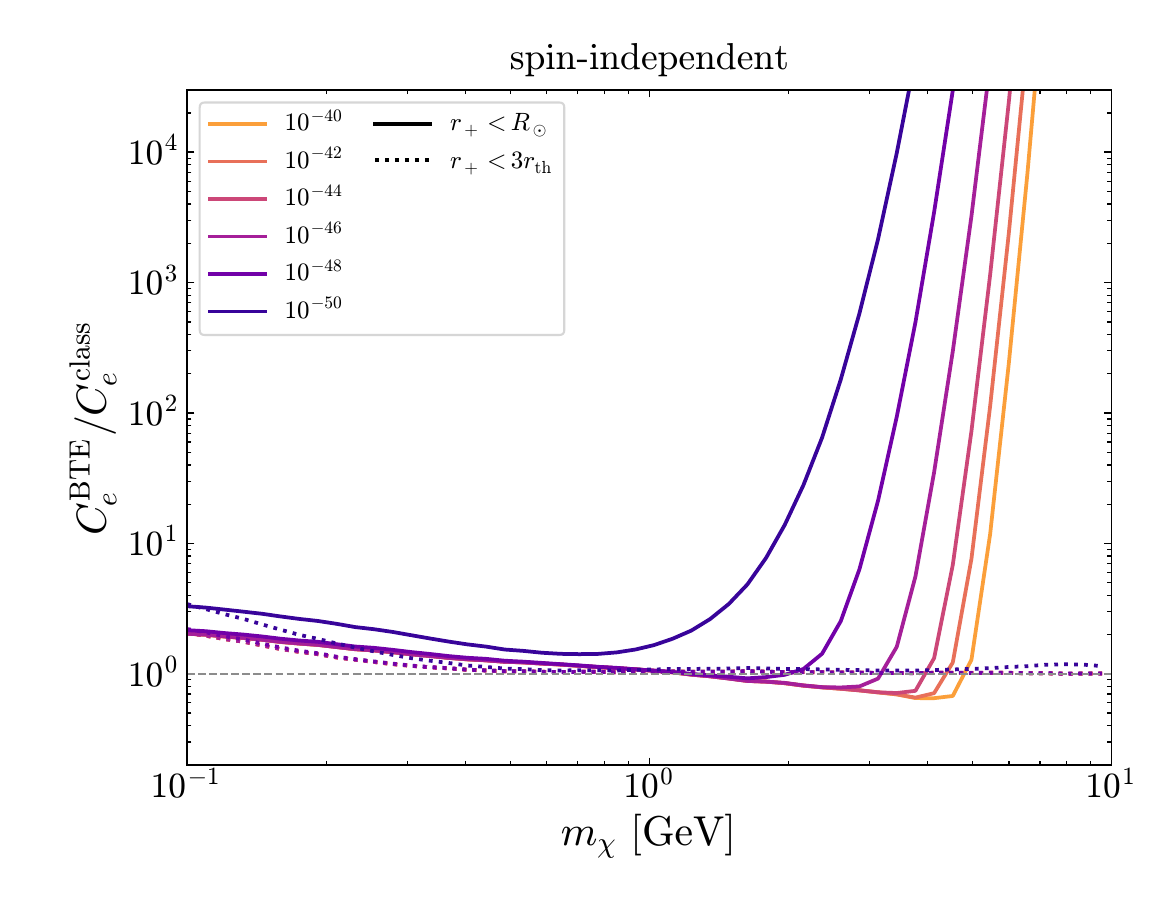}~
    \includegraphics[width=.48\textwidth]{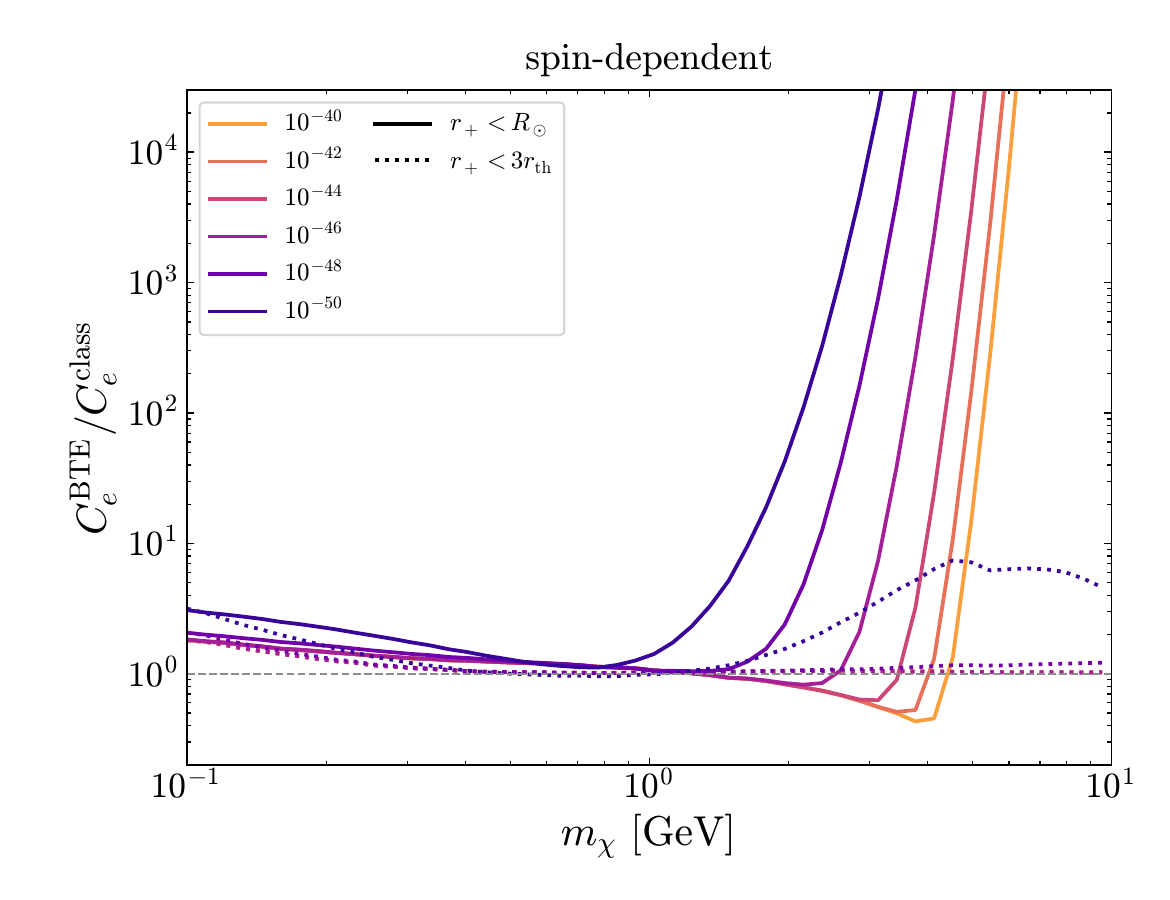}
    \caption{BTE-to-classical ratio of the evaporation coefficient at $t_\odot$,
    for SI (left) and SD (right) scattering.
    Colors indicate the different cross sections.
    Solid and dotted curves include only orbits with $r_+<R_\odot$ and
    $r_+<3r_{\rm th}$, respectively, where $r_{\rm th}$ is the thermal radius
    of the classical distribution. For each selection, the evaporation rate
    is divided by the number of selected particles in each treatment.}

\label{fig:C_e_ratio}
\end{figure*}

To compare the BTE with the classical treatment of Ref.~\cite{Garani:2017jcj}, in which frequent collisions with nuclei give an isothermal radial density $n_\chi^{\rm iso}(r)$ of captured DM, we evolve both from an empty population, $N_\chi=0$ at $t=0$, under continuous capture, annihilation and evaporation up to one solar age, $t_\odot=4.6~{\rm Gyr}$, with $\langle\sigma v\rangle=3\times10^{-26}~{\rm cm^3\,s^{-1}}$. In the BTE approach, we solve the master equation, Eq.~\eqref{eq:master_eq}, for $n_\chi(\xi,t)$ on the $192\times48$ grid in $(E,L)$ described in App.~\ref{app:grid} and evolve it with the implicit backward differentiation formula method,\footnote{We use the SciPy~\cite{Virtanen:2019joe} function \texttt{scipy.integrate.BDF} for numerical implementation.} with the kernels of Eqs.~\eqref{eq:Kcoll}, \eqref{eq:orbital-evaporation}, \eqref{eq:orbital-capture-source} and \eqref{eq:Kann}, and obtain $n_\chi(r,t)$ from Eq.~\eqref{eq:nx(r)_reconstruction}. In the classical treatment, we solve Eq.~\eqref{eq:evol_classical} for $N_\chi(t)$, computing $C_{c,a,e}$ as in Ref.~\cite{Garani:2017jcj}.

We adopt a local halo density of $\rho_\infty = n_\infty m_\chi=0.3~{\rm GeV\,cm^{-3}}$, and the DM halo speed $v_\infty$ in the solar frame follows a
Maxwellian
distribution~\cite{Garani:2017jcj},
\begin{align}
&f_\odot(v_\infty)
= \nonumber\\
&\quad \sqrt{\frac{3}{2\pi}}
\frac{v_\infty}{v_\odot v_d}
\left[
e^{-\frac{3(v_\infty-v_\odot)^2}{2v_d^2}}
-
e^{-\frac{3(v_\infty+v_\odot)^2}{2v_d^2}}
\right],
\label{eq:halo_vec_dist}
\end{align}
where $v_\odot=230~{\rm km\,s^{-1}}$ is the speed of the Sun relative to the halo
and $v_d=220~{\rm km\,s^{-1}}$ is the root-mean-square speed of halo DM in the Galactic frame.
The distribution is truncated at the Galactic escape speed of $544~{\rm km\,s^{-1}}$ in the Galactic frame~\cite{Smith:2006ym}, which removes only a small high-speed tail from Eq.~\eqref{eq:halo_vec_dist}.
We also neglect perturbations from the planets~\cite{Peter:2009mk}.\footnote{Planetary perturbations mainly affect DM on orbits that extend far beyond the Sun and would pull part of it away from the Sun, which only reduces the annihilation rate further~\cite{Peter:2009mk}.}

The parameter space considered here is $0.1\leq m_\chi/{\rm GeV}\leq10^3$ and
$10^{-50}\leq\sigma_{\chi p}/{\rm cm^2}\leq10^{-40}$, where $\sigma_{\chi p}$ is the DM-proton cross section, for both
SI and SD interactions.
For SI scattering, we include all 29 elements of the AGSS09 solar model and assume isospin symmetry, giving
$\sigma_{\chi A}=\sigma_{\chi p}A^2(\mu_{\chi A}/\mu_{\chi p})^2$,
where $\mu_{\chi A,p}$ are the DM-nucleus/nucleon reduced masses, and we use the Helm form factor
\begin{equation}
F_{A,{\rm SI}}(q)=\frac{3j_1(qR_A)}{qR_A}e^{-q^2s^2/2},
\end{equation}
where $j_1$ is the spherical Bessel function of the first kind.
The nuclear radius is $R_A=\sqrt{c_A^2+7\pi^2a^2/3-5s^2}$, with
$c_A=(1.23A^{1/3}-0.6)~{\rm fm}$, $a=0.52~{\rm fm}$,
$s=0.9~{\rm fm}$
and
$F_A(0)=1$.

For SD scattering, we take proton-only couplings, giving $\sigma_{\chi A}=\sigma_{\chi p}\langle S_{p,A}\rangle^2(\mu_{\chi A}/\mu_{\chi p})^2(4/3)(J_A+1)/J_A$, where $J_A$ is the nuclear spin and $\langle S_{p,A}\rangle$ is the expectation value of the proton spin in the nucleus, taken from \texttt{DarkSUSY}~6.4.3~\cite{Bringmann:2018lay}. The momentum dependence is described by the form factor
\begin{equation}
F_{A,{\rm SD}}^2(q) = \exp\left(-\frac{q^2R_A^2}{3}\right),
\end{equation}
where $R_A=\kappa_A(0.91A^{1/3}+0.30)~{\rm fm}$.
We include ${}^1{\rm H}$, ${}^3{\rm He}$, ${}^{13}{\rm C}$, ${}^{14}{\rm N}$, ${}^{15}{\rm N}$ and ${}^{17}{\rm O}$, treating hydrogen as a point particle with $F_{{\rm H,SD}}=1$ and using $\kappa_A=(1,1,1.04,1.04,1.12)$ for the other five nuclei in the order listed.\footnote{Our solar model resolves isotopes only up to oxygen, and these six are its spin-carrying isotopes, while \texttt{DarkSUSY} also includes heavier ones, such as ${}^{23}{\rm Na}$ and ${}^{27}{\rm Al}$. We estimate that adding them, with the neutron couplings of the Higgsino model in Sec.~\ref{sec:multicomponent} as well, changes $C_c$ by at most about 0.2\% in the parameter space we consider.} With these spin factors, ${}^3{\rm He}$, ${}^{13}{\rm C}$ and ${}^{17}{\rm O}$ carry their spin only in the unpaired neutron, so $\langle S_{p,A}\rangle=0$ and they do not contribute for proton-only couplings.
The thermal velocities of the nuclei
follow the Maxwellian distribution
\begin{equation}
f_A(r,u)=\left[\frac{m_A}{2\pi T_\odot(r)}\right]^{3/2}
 \exp\left[-\frac{m_Au^2}{2T_\odot(r)}\right],
 \label{eq:nuclear_velocity_input}
\end{equation}
where $m_A$ is the nuclear mass and $T_\odot(r)$ is the solar temperature at $r$.
The two treatments give the same capture rate $C_c$, while the BTE also keeps the orbital distribution of newly captured DM through $\Kcap$.

\subsection{Validity of the orbit average}
\label{sec:bte_validity}

The orbit-averaging approximation
requires $\tau_{\rm coll}\gg\tau_r$, so that the number of collisions per orbit, $N_{\rm coll}=\tau_r/\tau_{\rm coll}$, is small.
Fig.~\ref{fig:bte_validity} shows the cross sections below
which $N_{\rm coll}<0.1$ for 99\% of the bound DM at all times during the evolution.
Since the approximation fails only when $N_{\rm coll}$ approaches unity, this criterion leaves a margin of one order of magnitude, and even at the largest SI cross section we consider, $\sigma_{\chi p}=10^{-40}~{\rm cm^2}$, $N_{\rm coll}$ remains far below one.
At much larger cross sections, frequent collisions interrupt the orbital
motion, and DM diffuses in the short-mean-free-path regime studied in Refs.~\cite{Gould:1989hm,Leane:2022hkk}.

\subsection{Radial number density}
\label{sec:radial_profiles}

\begin{figure*}
    \centering
    \includegraphics[width=.48\textwidth]{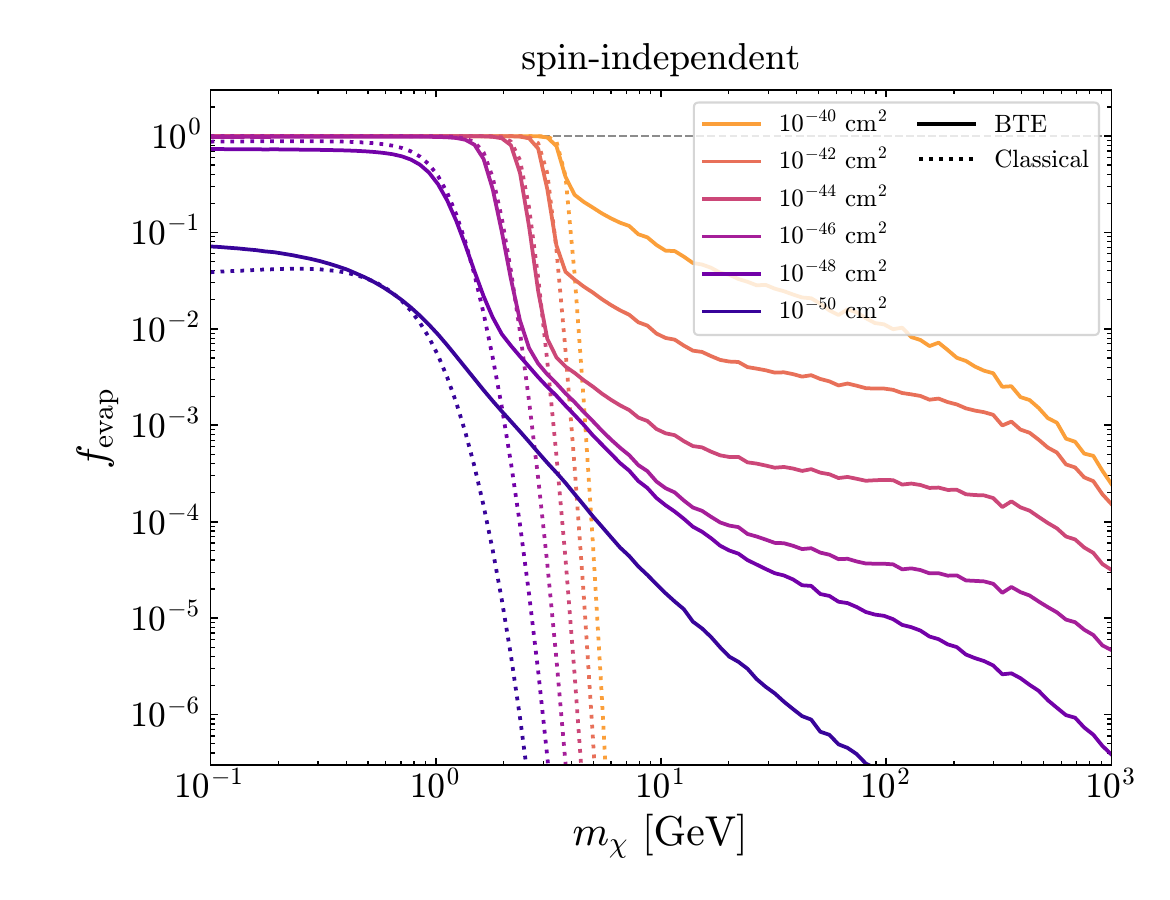}~
    \includegraphics[width=.48\textwidth]{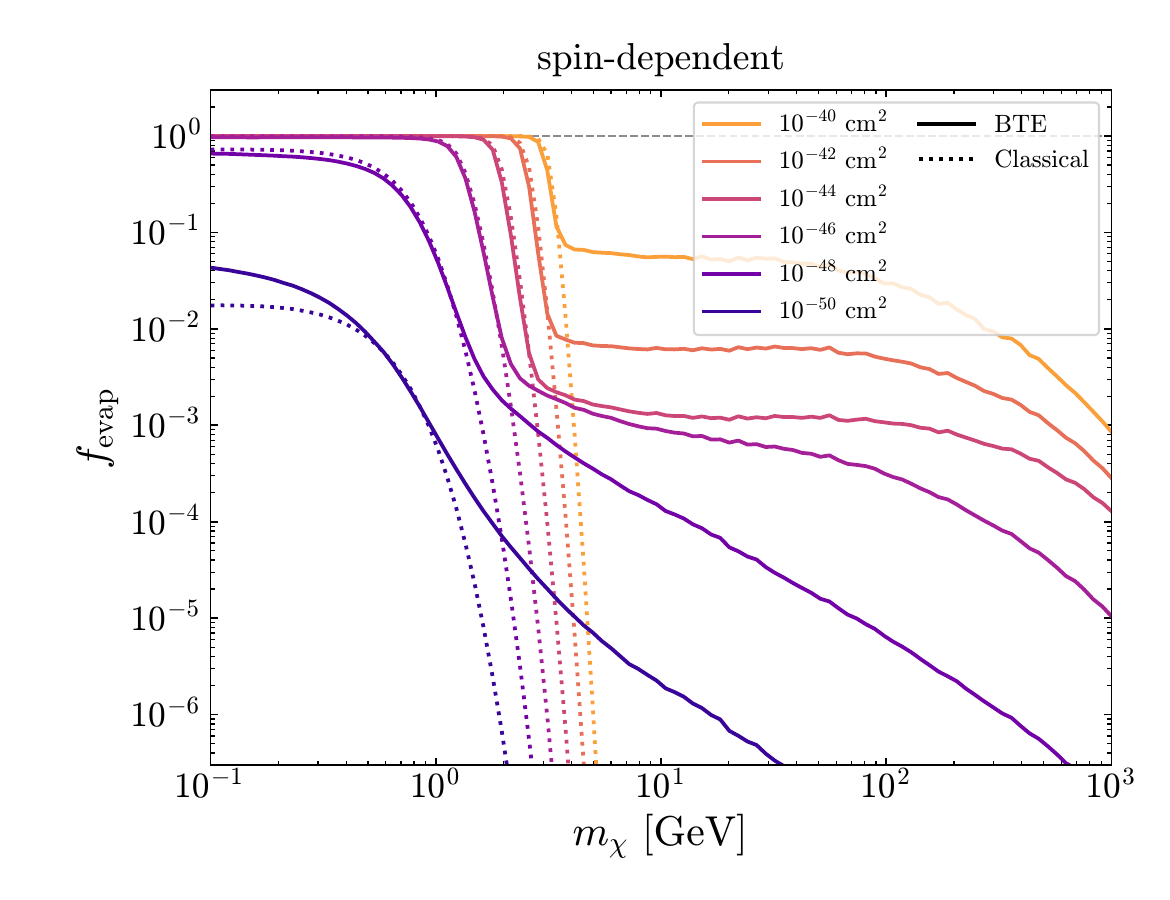}
    \caption{Accumulated evaporation fraction, Eq.~\eqref{eq:evaporation_fraction},
    at $t_\odot$ for SI (left) and SD (right) scattering.
    Solid and dotted curves show the BTE and classical results, respectively,
    with $\sigma_{\chi p}$ indicated by color.}

\label{fig:ratio_bound_evap}
\end{figure*}

Fig.~\ref{fig:nx_evolution} compares the radial densities for SD scattering
at $m_\chi=100~{\rm GeV}$, with
$\sigma_{\chi p}=10^{-40}$ and $10^{-50}~{\rm cm^2}$.
The profiles of newly captured DM (orange curves), set by the capture kernel alone, have the same shape for the two cross sections, while
their values
reflect the different capture rates.
The BTE profiles have tails that extend much farther than the classical isothermal profiles, because freshly captured particles are weakly bound ($E\to0$) and occupy large orbits, often
with apocenters beyond the solar surface.

For $\sigma_{\chi p}=10^{-40}~{\rm cm^2}$, repeated collisions ($\sim10^{9}$ per particle over $t_\odot$ on average) transfer
energy to solar nuclei and shrink the orbits, so that by one solar age the central
population approaches the classical isothermal distribution. A dilute tail remains
beyond the thermal radius, $r_{\rm th}\sim0.01R_\odot$ for this mass, because continuous
capture keeps supplying particles that have not yet thermalized.
For $\sigma_{\chi p}=10^{-50}~{\rm cm^2}$, a particle undergoes much less than one collision on average over the solar age, so it does not thermalize, as anticipated in Refs.~\cite{Peter:2009mk,Widmark:2017yvd}. The distribution then
preserves the broad shape set by capture as $N_\chi$ grows. At $t_\odot$, its central density is more than six orders of magnitude below the classical one, which suppresses the annihilation rate, as discussed next.

\subsection{Annihilation rate}\label{sec:annihilation_rate}

The annihilation rate follows from $\Kann$ given in Eq.~\eqref{eq:Kann},
\begin{equation}
\Gamma_A^{\rm BTE} = -\frac{1}{2} \int_{\Phi(0)}^0 dE \int_0^{L_\odot(E)} dL\,\Kann[n_\chi](\xi),\label{eq:Gamma_A}
\end{equation}
where the factor $1/2$ accounts for the two particles removed in each annihilation.
In evaluating $\Gamma_A^{\rm BTE}$, we restrict the radial integral in Eq.~\eqref{eq:Kann} to $r<R_\odot$, while the evolution itself includes annihilation at all radii.

Fig.~\ref{fig:Gamma_A_ratio} shows the ratio of $\Gamma_A^{\rm BTE}$ to the classical annihilation rate $\Gamma_A^{\rm class}$, obtained by assuming $n^{\rm iso}_\chi(r)$.
Above a few GeV, where evaporation is weak, the BTE annihilation rate is suppressed most strongly at small $\sigma_{\chi p}$ and large $m_\chi$.
In this regime, DM does not thermalize within the solar age and keeps the broad distribution set by capture, with most particles on orbits that extend well beyond the solar surface, as in the right panel of Fig.~\ref{fig:nx_evolution}.
This dilute distribution reduces $\Gamma_A$ through its dependence on $n_\chi^2$.
Annihilation outside the Sun, which is negligible for thermalized DM, would increase $\Gamma_A^{\rm BTE}$ in this regime by at most a factor of a few, but the distribution remains too dilute for this increase to overcome the suppression.

The suppression is much stronger for SD than for SI scattering at the same $\sigma_{\chi p}$, which reflects how efficiently collisions transfer energy.
For SI scattering, the ratio stays roughly between 0.5 and 3 above about 1~GeV for $\sigma_{\chi p}\geq10^{-48}~{\rm cm^2}$, because collisions with nuclei are both frequent and efficient in transferring energy.
Energy exchange in an elastic collision is most efficient for comparable DM and nuclear masses, with a mean energy loss fraction of $\eta =2m_\chi m_A/(m_\chi+m_A)^2$, which favors helium, oxygen and iron.
Helium and oxygen are among the most abundant elements in the Sun, and the coherent enhancement $A^2(\mu_{\chi A}/\mu_{\chi p})^2$ of SI scattering compensates for the lower abundance of iron.
At $\sigma_{\chi p}=10^{-50}~{\rm cm^2}$, collisions become too rare to thermalize DM within the solar age, although the suppression remains weakest where these nuclei are most effective, which gives the broad bump from a few to several tens of GeV.
In contrast, SD scattering has no coherent $A^2$ enhancement and is dominated by hydrogen
for the interactions considered here. Energy exchange becomes inefficient
when $m_\chi$ is much larger than the proton mass, so the suppression grows rapidly with increasing $m_\chi$.

Fig.~\ref{fig:Gamma_A_ratio_scan} shows $\Gamma_A^{\rm BTE}/\Gamma_A^{\rm class}$ over the $m_\chi$-$\sigma_{\chi p}$ plane for SI and SD interactions.
In the region with $\Gamma_A^{\rm BTE}/\Gamma_A^{\rm class}<0.5$, i.e., where the classical treatment overestimates the annihilation rate by more than a factor of two, the evolution of the distribution has to be followed with the BTE.
Below a few GeV and at sufficiently large cross sections, evaporation becomes dominant and determines the captured population, as discussed next.

\subsection{Evaporation}\label{sec:evap_mass}

To determine the DM mass at which evaporation becomes substantial, we use the evaporation mass $m_{\rm evap}$, following the definition in Refs.~\cite{Garani:2017jcj,Garani:2021feo},
\begin{equation}
 \left.-\frac{C_e(t)N_\chi(t)}{C_c}\right|_{m_\chi=m_{\rm evap}}
  =\frac{10}{11},
 \label{eq:evaporation_mass}
\end{equation}
where $-C_e$ is the evaporation rate per particle.
The criterion therefore selects the mass at which the total evaporation rate is
$10/11$ of $C_c$ at $t_\odot$, without assuming number equilibrium between capture and evaporation, $\dot{N}_\chi=0$.\footnote{When annihilation is negligible, number equilibrium between capture and evaporation, $\dot{N}_\chi=0$, gives $N_\chi^{\rm eq}=-C_c/C_e$. The definition of $m_{\rm evap}$ in Refs.~\cite{Garani:2017jcj,Garani:2021feo} requires the difference between $N_\chi$ and $N_\chi^{\rm eq}$ to be 10\% of $N_\chi$, $|N_\chi - N_\chi^{\rm eq}|=N_\chi/10$. Since $N_\chi < N_\chi^{\rm eq}$ before number equilibrium is reached, this gives the factor $10/11$.}
Fig.~\ref{fig:m_evap} shows the BTE (solid red) and classical (solid green)
results for $m_{\rm evap}$, while the green dashed curve assumes $\dot{N}_\chi=0$.
At $\sigma_{\chi p}=10^{-44}~{\rm cm^2}$, the largest cross section in Fig.~\ref{fig:m_evap}, the classical values agree with those of Ref.~\cite{Garani:2017jcj} for both SI and SD scattering.
Since $m_{\rm evap}$ lies near the masses of hydrogen and helium, the most abundant nuclei in the Sun, $\eta$ is close to its maximum $1/2$ at $m_\chi\sim m_{\rm evap}$, and DM thermalizes easily there once collisions are frequent enough.
The BTE and classical values of $m_{\rm evap}$ thus agree for both SI and SD scattering down to $\sigma_{\chi p}\approx10^{-47}~{\rm cm^2}$.
At smaller cross sections, infrequent
collisions keep $n_\chi(r)$ away from an isothermal distribution, and less DM resides in the dense, hot core where most collisions occur.
As a result, evaporation is less efficient, and the BTE value of $m_{\rm evap}$ falls significantly below
the classical one.

At fixed $\sigma_{\chi p}$, reducing $m_\chi$ far below the hydrogen mass gives
$\eta\sim 2m_\chi/m_A\ll1$.
With such inefficient energy exchange, the nonthermal distribution contains
more weakly bound particles near the solar surface,
which evaporate more easily.
Since the ratio of the BTE evaporation coefficient, $C_e^{\rm BTE}$, to the classical value, $C_e^{\rm class}$, is sensitive to the weakly bound tail, Fig.~\ref{fig:C_e_ratio} shows it only for the bound DM on orbits with $r_+<R_\odot$ (solid) or $r_+<3r_{\rm th}$ (dotted), for different cross sections.\footnote{The classical isothermal distribution, $f\propto n_\chi^{\rm iso}(r)\,e^{-m_\chi v^2/2T_\chi}\propto e^{-m_\chi E/T_\chi}$, depends only on $E$ and is isotropic in velocity~\cite{Gould:1987ju,BinneyTremaine}. Each particle therefore lies on an orbit with $L=rv\sin\alpha$ and a uniform distribution of $\cos\alpha$, and in the classical approach we apply the same $r_+$ selection as for the BTE.}
This ratio refers only to the selected orbits, whereas $m_{\rm evap}$ depends on the total evaporation rate of all bound DM, much of which lies on orbits beyond the Sun that rarely scatter, so the BTE can give a lower $m_{\rm evap}$ even where the ratio exceeds unity.
For $m_\chi\lesssim m_{\rm evap}\sim1$~GeV, more of the weakly bound particles on the selected orbits evaporate than in the classical treatment, which lowers the density close to the solar surface, at $r\sim0.3$--$1R_\odot$, and can also explain the suppression of
the annihilation rate at $m_\chi\lesssim1$~GeV shown in
Fig.~\ref{fig:Gamma_A_ratio}.
When only orbits with
$r_+<3r_{\rm th}$ are included (dotted curves), the two treatments give similar $n_\chi$,
and the ratio remains close to unity between about 0.5~GeV and a few GeV.

Once $m_\chi\gtrsim$ a few GeV, evaporation becomes negligible in the classical treatment, while the BTE keeps a weakly bound nonthermal tail that can still evaporate.
For $r_+<R_\odot$, this tail extends to the solar surface, and $C_e^{\rm BTE}/C_e^{\rm class}$ rises rapidly and diverges as $C_e^{\rm class}\to0$
(see App.~\ref{app:nx_tail} for the $n_\chi$ tail discussion).
For SD scattering with $\sigma_{\chi p}=10^{-50}~{\rm cm^2}$, inefficient collisions leave such a tail even within $3r_{\rm th}$, so the dotted curve also reaches $\mathcal{O}(10)$ around $m_\chi\sim5$~GeV.
However, such a divergence does not mean that evaporation still dominates capture well above $m_{\rm evap}$ in the BTE approach. Fig.~\ref{fig:ratio_bound_evap} shows the evaporated fraction,
\begin{equation}
 f_{\rm evap}=
  \frac{N_{\rm evap}}{N_{\rm evap}+N_{\rm bound}},
 \label{eq:evaporation_fraction}
\end{equation}
where $N_{\rm evap}$ is the number of evaporated particles and
$N_{\rm bound}$ is the number of particles that remain bound.
The fraction drops significantly at $m_\chi\gg m_{\rm evap}$, showing that evaporation has little impact on the captured DM there.

\section{Generalization to two-component DM}
\label{sec:multicomponent}

\begin{table}[t]
\caption{Higgsino benchmark parameters~\cite{Wu:2026nhi}.
Proton ($p$) and neutron ($n$) couplings differ in both SI and SD scattering.
Here $\langle\sigma v\rangle$ refers to $\chi_1\chi_1$ annihilation, and
$\tau_2$ is the excited-state lifetime.}

\label{tab:higgsino_benchmark}
\begin{ruledtabular}
\begin{tabular}{lc}
Particle property& Value \\
\hline
$m_{\chi_1}$ & $1.1\times10^3~{\rm GeV}$ \\
$\sigma_p^{{\rm SI,inelastic}}$ & $6.8\times10^{-41}~{\rm cm^2}$ \\
$\sigma_n^{{\rm SI,inelastic}}$ & $7.6\times10^{-39}~{\rm cm^2}$ \\
$\sigma_p^{{\rm SI,elastic}}$ & $2.7\times10^{-49}~{\rm cm^2}$ \\
$\sigma_n^{{\rm SI,elastic}}$ & $2.8\times10^{-49}~{\rm cm^2}$ \\
$\sigma_p^{{\rm SD,elastic}}$ & $1.0\times10^{-45}~{\rm cm^2}$ \\
$\sigma_n^{{\rm SD,elastic}}$ & $7.7\times10^{-46}~{\rm cm^2}$ \\
$\langle\sigma v\rangle$ & $8.1\times10^{-27}~{\rm cm^3\,s^{-1}}$ \\

$\tau_2(\delta=370~{\rm keV})$ & $0.026~{\rm s}$ \\
\end{tabular}
\end{ruledtabular}
\end{table}

As an example of DM with several components, we take the Higgsino model~\cite{Wu:2026nhi,DiMauro:2026ldr}, one of the inelastic DM interpretations~\cite{Wu:2026nhi,DiMauro:2026ldr,deLima:2026shq,Fan:2026hzw,Okada:2026upm} of the high-recoil candidate reported by LZ~\cite{LZ:2026axp}, for which solar-neutrino searches provide complementary tests~\cite{Pospelov:2026ewn,DiMauro:2026dqp,Bose:2026ndd,DiMauro:2026ymt,Bose:2026szs}. The two neutral states, with masses $m_{\chi_2}=m_{\chi_1}+\delta$, interact through inelastic scattering $\chi_1A\leftrightarrow\chi_2A$, elastic SI and SD scattering, radiative decay $\chi_2\to\chi_1+\gamma$ and annihilation $\chi_1\chi_1\to W^+W^-/ZZ$, with the cross sections and other model parameters~\cite{Wu:2026nhi} listed in Tab.~\ref{tab:higgsino_benchmark}.\footnote{We build $\sigma_{\chi A}$ as in Sec.~\ref{sec:numerical_results}, now with separate proton and neutron amplitudes given by $\sigma_{\chi p}$ and $\sigma_{\chi n}$ of Tab.~\ref{tab:higgsino_benchmark}. For SI scattering, $A^2$ is replaced by $(Z\pm(A-Z)\sqrt{\sigma_{\chi n}/\sigma_{\chi p}})^2$, where $Z$ is the atomic number here. For SD scattering, $\langle S_{p,A}\rangle^2$ is replaced by $(\langle S_{p,A}\rangle\pm\sqrt{\sigma_{\chi n}/\sigma_{\chi p}}\langle S_{n,A}\rangle)^2$, where $\langle S_{n,A}\rangle$ is the expectation value of the neutron spin. The relative sign is negative for $Z$-boson exchange (inelastic SI and elastic SD) and positive for Higgs-boson exchange (elastic SI).}

\subsection{Coupled master equation}\label{sec:coupled_master}

The halo contains a negligible abundance of $\chi_2$, because the excited state decays promptly through $\chi_2\to\chi_1+\gamma$, with a lifetime $\tau_2(\delta)\simeq0.026~{\rm s}\,(370~{\rm keV}/\delta)^3$~\cite{Krall:2017xij}.
Halo $\chi_1$ is captured mainly through the inelastic scattering $\chi_1A\to\chi_2A$ on solar nuclei, because its elastic cross sections are much smaller than the inelastic ones (see Tab.~\ref{tab:higgsino_benchmark}), and the $\chi_2$ produced in this way decays back to $\chi_1$ before it can scatter again.
The populations of $\chi_1$ and $\chi_2$ on orbit $\xi$ then evolve according to the coupled master equations
\begin{widetext}
\begin{subequations}\label{eq:coupled_master}
\begin{align}
\frac{\partial n_{\chi_1}(\xi)}{\partial t} &= \Kcoll{}_{,11}[n_{\chi_1}](\xi) + \Kcoll{}_{,21}[n_{\chi_2}](\xi)
 + \Kevap{}_{,1}[n_{\chi_1}](\xi)+\Kann{}_{,1}[n_{\chi_1}](\xi) + \frac{n_{\chi_2}(\xi)}{\tau_2}, \\
\frac{\partial n_{\chi_2}(\xi)}{\partial t} &= \Kcap{}_{,2}[n_{\chi_2}](\xi) +\Kcoll{}_{,12}[n_{\chi_1}](\xi) + \Kcoll{}_{,22}[n_{\chi_2}](\xi)+ \Kevap{}_{,2}[n_{\chi_2}](\xi)
- \frac{n_{\chi_2}(\xi)}{\tau_2},\label{eq:coupled_nx2}
\end{align}
\end{subequations}
\end{widetext}
where, for $i\ne j$, $\mathcal K_{{\rm coll},ij}$ describes the production of bound
$\chi_j$ from $\chi_i$, while $\mathcal K_{{\rm coll},ii}$ includes
elastic redistribution and the loss through conversion into the other state.
The last terms of the two equations describe the gain of $\chi_1$ and the loss of $\chi_2$ through the decay $\chi_2\to\chi_1+\gamma$.
The capture of halo $\chi_1$ into bound $\chi_2$ is described by $\mathcal{K}_{{\rm cap},2}$ in Eq.~\eqref{eq:coupled_nx2}, and we omit the small contributions from halo $\chi_2$ and from the elastic capture of $\chi_1$ from Eq.~\eqref{eq:coupled_master}.

The photon emitted in the decay carries energy $E_\gamma\simeq\delta$ in the nonrelativistic limit, so
$\chi_1$ keeps nearly the same orbital energy as $\chi_2$,
$E_{\chi_1}\simeq E_{\chi_2}$, and receives only a small recoil momentum
$|\Delta {\bf p}|\sim\delta$.
The fractional change in angular momentum is then $|\Delta L|/L \sim |\bfr \times \Delta{\bf p}|/(r m_{\chi_1}v_t)\sim \delta/(m_{\chi_1}v_t)$,
which is about $10^{-4}$ for $v_t\sim10^3~{\rm km\,s^{-1}}$.
Our numerical calculation shows that this recoil has little impact, so Eq.~\eqref{eq:coupled_master} neglects it and places the daughter $\chi_1$ on the orbit of its parent $\chi_2$.

Since $\chi_2$ decays promptly, $n_{\chi_2}$ stays negligible, and the bound population consists of $\chi_1$, whose annihilation rate is
\begin{equation}
\Gamma_A = -\frac{1}{2} \int_{\Phi(0)}^0 dE \int_0^{L_\odot(E)} dL\,\Kann{}_{,1}[n_{\chi_1}](\xi).
\label{eq:Gamma_A_chi1}
\end{equation}
Our calculation includes the excitation of bound $\chi_1$, described by $\mathcal K_{{\rm coll},12}$.
It does not make the captured DM convert back and forth many times between $\chi_1$ and $\chi_2$, because after three to four excitations $\chi_1$ has lost enough kinetic energy that the energy available in collisions with thermal nuclei no longer exceeds $\delta$.

\subsection{Constraint}

\begin{figure}
 \centering
 \includegraphics[width=.9\columnwidth]{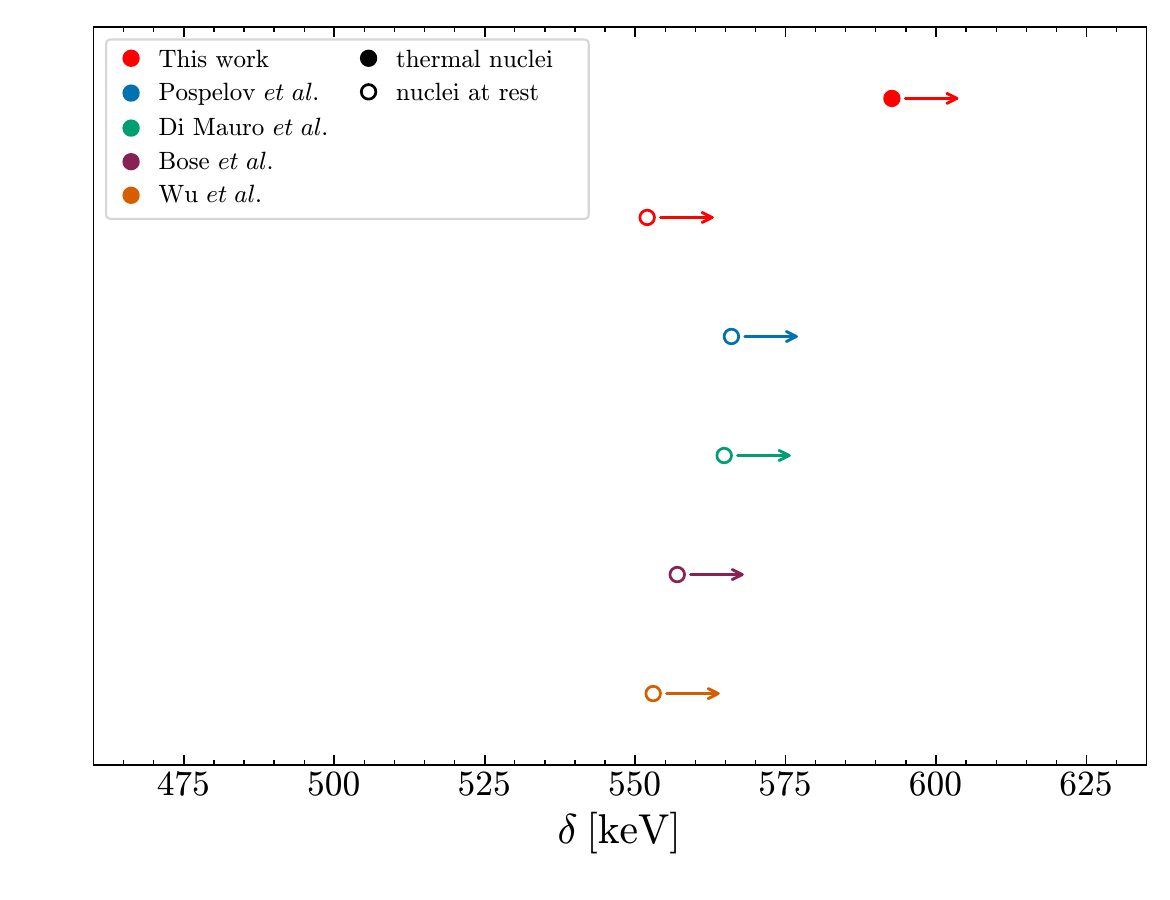}
 \caption{Lower limits on $\delta$ for the Higgsino model. Filled and open red circles show our
 results with thermal nuclei and nuclei at rest, respectively, during capture.
 Right-pointing arrows mark the allowed region. Limits from Refs.~\cite{Pospelov:2026ewn,DiMauro:2026dqp,Bose:2026ndd,Wu:2026nhi}
 are shown for comparison.}

\label{fig:delta_constraint}
\end{figure}

\begin{figure}
 \centering
 \includegraphics[width=.96\columnwidth]{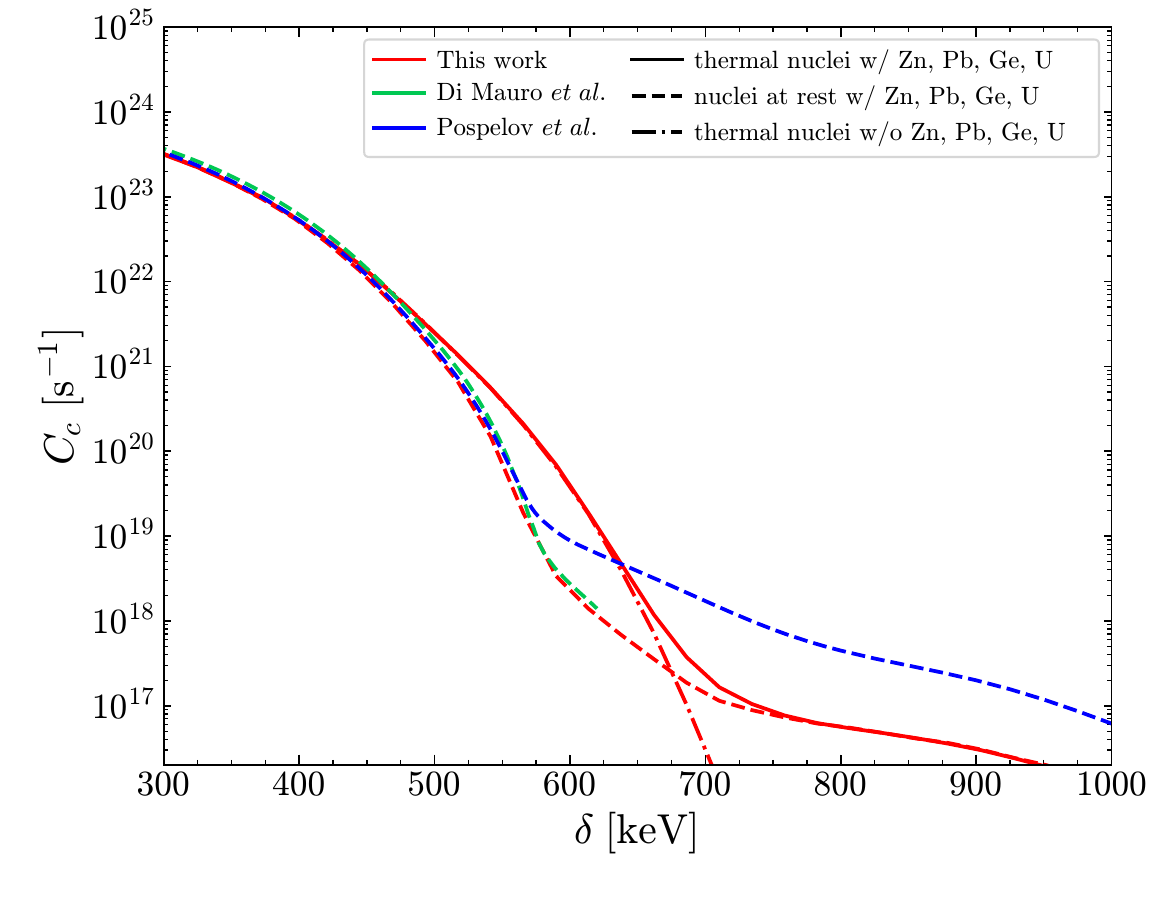}
 \caption{Inelastic $C_c$ vs.\ $\delta$.
 Our red solid and dashed curves use 33 nuclei, the 29 AGSS09 elements plus Zn, Ge, Pb and U, with and without thermal motion, respectively. The red dot-dashed curve includes thermal motion but omits Zn, Ge, Pb and U.
 Green and blue dashed curves show results for nuclei at rest from Refs.~\cite{DiMauro:2026dqp,Pospelov:2026ewn} with heavy elements included.
   }

\label{fig:higgsino_capture}
\end{figure}

To constrain $\delta$, we evolve Eq.~\eqref{eq:coupled_master} from $n_{\chi_1}=n_{\chi_2}=0$ at $t=0$ under continuous capture up to $t_\odot$, evaluate $\Gamma_A$ at the present day with Eq.~\eqref{eq:Gamma_A_chi1}, and compare it with the upper limit on the total annihilation rate, $\Gamma_A^{\rm IC}=2.9\times10^{19}~{\rm s^{-1}}$, set by the 10-yr IceCube limit on the $W^+W^-$ channel~\cite{IceCube:2025fcu}. The $W^+W^-$ channel, which we use alone for the neutrino signal, has a branching ratio of about $0.54$ for this benchmark, while $ZZ$ has $0.46$. Here we use the same numerical setup as in Sec.~\ref{sec:numerical_results}, generalized to the two components, with the 29 AGSS09 elements as targets.
Fig.~\ref{fig:delta_constraint} shows our lower limit on $\delta$ (red) together with those of Refs.~\cite{Pospelov:2026ewn,DiMauro:2026dqp,Bose:2026ndd,Wu:2026nhi}, where the right-pointing arrows mark the allowed region.
Our calculation excludes $\delta\lesssim592$~keV when the thermal motion of nuclei is included in the inelastic capture rate (filled circle) and $\delta\lesssim552$~keV when it is neglected (open circle). Adding Zn, Ge, Pb and U changes both circles by only about 0.1~keV.

The constraints of Refs.~\cite{Pospelov:2026ewn,DiMauro:2026dqp,Bose:2026ndd,Wu:2026nhi} use the classical number equation with $2\Gamma_A\simeq C_c$, which our evolution supports, but assume nuclei at rest in the capture rate. With nuclei at rest, our constraint weakens by about 40~keV (open circle) and becomes comparable to theirs.
We show below that the thermal motion of the nuclei allows capture at larger $\delta$ even without the additional heavy elements, such as Zn, Ge, Pb and U, that these studies include~\cite{Pospelov:2026ewn,DiMauro:2026dqp,Bose:2026ndd,Wu:2026nhi}.

Fig.~\ref{fig:higgsino_capture} exhibits our inelastic $C_c$ as a function of $\delta$ for different treatments of the nuclei, with the results of Refs.~\cite{DiMauro:2026dqp,Pospelov:2026ewn} for reference.
With nuclei at rest, our $C_c$ (red dashed curve) agrees with that of Ref.~\cite{DiMauro:2026dqp}.
Near our constraint and with thermal motion included, the additional heavy elements barely change $C_c$ (red solid and dot-dashed curves).
Without thermal motion (red dashed curve), $C_c$ is much lower for $500~{\rm keV}\lesssim\delta\lesssim700~{\rm keV}$, because inelastic capture requires the center-of-mass (CoM) energy of the DM-nucleus pair,
\begin{equation}
E_{k,{\rm cm}}=\frac{1}{2}\mu_{\chi A}g^2\stackrel{u\to0}{=}\frac{1}{2}\mu_{\chi A} w^2,
\end{equation}
to exceed $\delta$, where $g=|{\bf w}-{\bf u}|$ is the relative speed, and the rightmost result is the limit of nuclei at rest.
For a nucleus at rest, all the energy comes from the incident DM, so capture requires $w>\sqrt{2\delta/\mu_{\chi A}}$, and heavier targets allow capture at larger $\delta$.

For example, at $r=0.01R_\odot$, $w\sim 1410~{\rm km\,s^{-1}}$ for a mean halo speed $\bar{v}_\infty\sim300~{\rm km\,s^{-1}}$, and the largest $\delta$ accessible in scattering on iron at rest is about 550~keV.
With thermal motion, a head-on collision with an iron nucleus moving at the mean thermal speed
$\bar{u}\approx 77~{\rm km\,s^{-1}}$ could allow capture up to $\delta\sim 610$~keV. For isotropically distributed incident
directions, $E_{k,{\rm cm}}>\delta$ holds for about half of the directions at $\delta=550$~keV and for about 10\% at $\delta=600$~keV. The thermal motion of the nuclei alone can therefore raise the $\delta$ threshold for inelastic capture without the four heavy elements.
Ignoring the thermal motion of the nuclei in $C_c$ is also inconsistent with the classical isothermal treatment, in which the DM temperature $T_\chi$ results from energy exchange with the same thermal nuclei and is thus closely tied to the solar temperature. With solar nuclei at rest, collisions would eventually drive $T_\chi\to0$, and $n_\chi$ would collapse into the center of the Sun.

\begin{figure*}[t]
    \centering
    \includegraphics[width=.8\textwidth]{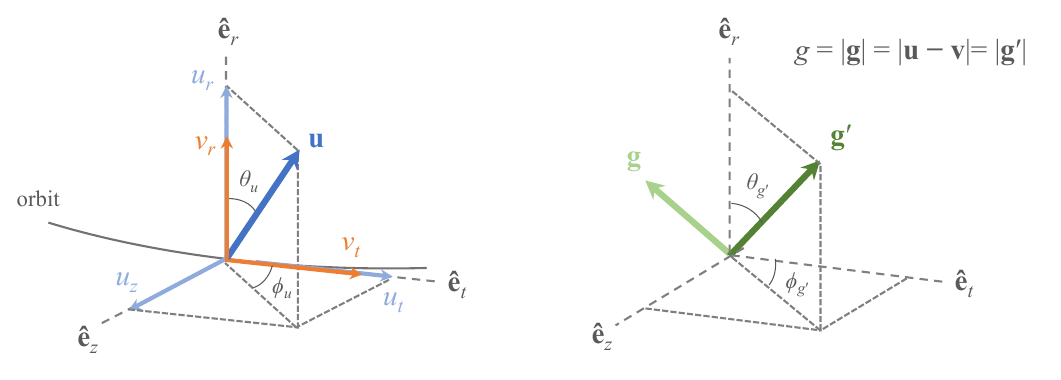}

    \caption{Collision kinematics. Left: the incident DM velocity
    has radial and tangential components $v_r$ and $v_t$, and the thermal
    velocity ${\bf u}$ of the nucleus has polar angle $\theta_u$ and
    azimuthal angle $\phi_u$. Right: elastic scattering rotates
    the relative velocity from ${\bf g}$ to ${\bf g}'$, specified by
    $\theta_{g'}$ and $\phi_{g'}$, while preserving its magnitude.}
    \label{fig:kinematics}
\end{figure*}

\section{Summary}
\label{sec:summary}

We have followed the evolution of DM captured by stars with the BTE approach, without assuming, as most studies do, that the captured population has already thermalized with the stellar nuclei.
When a bound particle completes many orbits between collisions, the distribution can be averaged over the orbital motion, which leads to a master equation
that evolves capture, scattering between bound orbits, annihilation and evaporation simultaneously.
Integrating it over orbits recovers the classical number evolution equation, with the annihilation and evaporation coefficients set by the evolving distribution.

For DM captured by the Sun, the BTE reproduces the classical evaporation mass and, above a few GeV, the classical annihilation rate when scattering on nuclei is frequent, which supports the classical treatment in this regime.
We find that the classical treatment breaks down when this scattering is rare or exchanges little energy per collision.
The captured DM then does not thermalize within the solar age and keeps the broad distribution it had at capture, with many particles on orbits that extend far beyond the solar surface.
At $m_\chi=100$~GeV, the classical treatment overestimates the annihilation rate inside the Sun by more than a factor of two for $\sigma_{\chi p}$ below about $10^{-46}~{\rm cm^2}$ for SD and $10^{-49}~{\rm cm^2}$ for SI scattering, and the suppression reaches many orders of magnitude at larger $m_\chi$ and smaller $\sigma_{\chi p}$.
With less DM in the hot core, evaporation is also weaker, and the evaporation mass falls below the classical value.

As an example of DM with several components, we applied the BTE to the inelastic Higgsino model, a possible interpretation of the recent LZ result, and recomputed the constraint on the mass splitting $\delta$.
The full evolution supports the relation $2\Gamma_A\simeq C_c$ assumed in recent studies~\cite{Pospelov:2026ewn,DiMauro:2026dqp,Bose:2026ndd,Wu:2026nhi}, and the IceCube limit excludes $\delta\lesssim592$~keV.
The constraint is stronger by tens of keV than in these studies, mainly because the thermal motion of solar nuclei in the capture rate allows inelastic capture at larger $\delta$ without the inclusion of additional heavy elements.

The solar results show that, when scattering on nuclei is weak, the signals of captured DM should be computed from its evolving distribution rather than from an assumed thermal one.
The BTE approach applies whenever bound DM completes many orbits before its orbit changes appreciably, a condition that is not specific to the Sun.
One natural extension is DM self-interaction, which adds collisions among the bound particles that compete with scattering on nuclei in setting their distribution~\cite{Zentner:2009is,Chen:2015poa}.
Heavy DM changes its orbit only slightly in each collision, so the same approach can follow its capture and thermalization in WDs, NSs and massive stars~\cite{Bell:2024qmj,Bell:2023ysh,Robles:2025dbd} once the collision kernels are adapted to these environments.
Coupling the BTE with stellar evolution would further replace the static star assumed here with a changing one.
With these extensions, the BTE approach could provide the evolving distribution of captured DM, and hence its signals, in the other stellar objects used to search for DM.

\begin{acknowledgments}
Y.-H.\ L. and M.-R.\ W. acknowledge support of the National Science and Technology Council, Taiwan under Grants No.~115-2112-M-001-037-MY3, No.~111-2628-M-001-003-MY4, No.~115-2112-M-001-054-MY5, and Academia Sinica under Project No.~AS-IV-114-M04.
Y.-H.\ L. further acknowledges support from the Shui-Chin Lee Foundation through the Shui-Chin Lee Fellowship.
M.-R.\ W. also acknowledges support of the Physics Division of the National Center for Theoretical Sciences, Taiwan.
S.\ Y. is supported by IBS under the Project code No.\ IBS-R018-D1
and the Basic Science Research Program through the NRF funded by the Ministry of Education (No.~RS-2026-25571203).
We acknowledge the computing resources provided by the Academia Sinica Grid-computing Center (ASGC). We also acknowledge the use of Anthropic Claude and OpenAI Codex to optimize the code and manage parallel computing at ASGC.

\end{acknowledgments}

\appendix

\section{Scattering kinematics}
\label{app:kinematics}

A collision changes the DM velocity at a fixed position, and the outgoing energy and angular momentum follow from the new velocity.
At the collision point, we use an orthonormal basis $(\hat{\bf e}_r,\hat{\bf e}_t,\hat{\bf e}_z)$, where $\hat{\bf e}_r$ points radially outward, $\hat{\bf e}_t$ points along the tangential component of the incident DM velocity ${\bf v}$, and $\hat{\bf e}_z$ is perpendicular to both.
The velocities of the incident DM and the nucleus are then
\begin{equation}
  {\bf v}=\begin{pmatrix}v_r\\v_t\\0\end{pmatrix},\qquad
  {\bf u}=u\begin{pmatrix}
    \cos\theta_u\\\sin\theta_u\cos\phi_u\\
    \sin\theta_u\sin\phi_u
  \end{pmatrix},
  \label{eq:incoming-components}
\end{equation}
as shown in the left panel of Fig.~\ref{fig:kinematics}, and $d\Omega_u=d\cos\theta_u d\phi_u$ is the solid angle of ${\bf u}$.
In a spherically symmetric star, the results do not depend on the azimuth $\alpha$ of the tangential velocity about $\hat{\bf e}_r$, and this choice of $\hat{\bf e}_t$ sets $\alpha=0$.

The relative and CoM velocities are
\begin{equation}
  {\bf g}={\bf v}-{\bf u},\qquad
  {\bf V}=\frac{m_\chi{\bf v}+m_A{\bf u}}{m_\chi+m_A},
  \label{eq:relative-cm-velocity}
\end{equation}
with elastic scattering leaving ${\bf V}$ and the relative speed
$g=|{\bf g}|$ unchanged while rotating ${\bf g}$ to
\begin{equation}
  {\bf g}'=g\begin{pmatrix}
    \cos\theta_{g'}\\\sin\theta_{g'}\cos\phi_{g'}\\
    \sin\theta_{g'}\sin\phi_{g'}
  \end{pmatrix},
  \label{eq:outgoing-direction}
\end{equation}
where $d\Omega_{g'}=d\cos\theta_{g'}d\phi_{g'}$ is the solid angle of the outgoing relative velocity.
The outgoing DM velocity is
\begin{equation}
  {\bf v}'_f={\bf V}+\frac{m_A}{m_\chi+m_A}{\bf g}',
  \label{eq:outgoing-velocities}
\end{equation}
so the incident velocities and the outgoing direction determine $E_f'$ and $L_f'$,
\begin{equation}
  E_f'=\frac{v_f'^2}{2}+\Phi(r),\qquad
  L_f'=r\sqrt{v_{f,t}'^2+v_{f,z}'^2}.
  \label{eq:outgoing-orbit}
\end{equation}
Both components perpendicular to $\hat{\bf e}_r$ contribute
to $L_f'$, so scattering can change the orbital plane as well as
the orbital energy.
The momentum transfer is ${\bf q}=m_\chi({\bf v}_f'-{\bf v})$,
whose squared magnitude is
\begin{equation}
  q^2=2\mu_{\chi A}^2
  g^2(1-\cos\theta_{\rm cm}),
  \label{eq:momentum-transfer}
\end{equation}
where $\theta_{\rm cm}$ is the angle between ${\bf g}$ and ${\bf g}'$,
with $\cos\theta_{\rm cm}={\bf g}\cdot{\bf g}'/g^2$ for $g>0$.
This momentum transfer enters the nuclear form factor $F_A(q)$.

For inelastic scattering, $\chi_1A\to\chi_2A$, we neglect the mass splitting in the inertial masses, and energy conservation in the CoM frame gives
\begin{equation}
  g'^2=g^2-\frac{2\delta}{\mu_{\chi A}},
  \label{eq:inelastic-g}
\end{equation}
which requires $\mu_{\chi A}g^2/2>\delta$, while $g'^2=g^2+2\delta/\mu_{\chi A}$ for the reverse process.
Eq.~\eqref{eq:outgoing-velocities} still gives the outgoing DM velocity, now with $|{\bf g}'|=g'$, and the squared momentum transfer becomes
\begin{equation}
  q^2=\mu_{\chi A}^2\left(g^2+g'^2-2gg'\cos\theta_{\rm cm}\right),
  \label{eq:momentum-transfer-inelastic}
\end{equation}
with $\cos\theta_{\rm cm}={\bf g}\cdot{\bf g}'/(gg')$.
For $g'=g$, the above relation reduces to Eq.~\eqref{eq:momentum-transfer}.
Since the final-state phase space scales with $g'$, the factor $g$ in Eqs.~\eqref{eq:R_v2v'} and \eqref{eq:rate_K} becomes $g'$ for inelastic scattering, and the rate vanishes at threshold $\delta=\mu_{\chi A}g^2/2$.

\section{Orbits that cross the solar interior}
\label{app:solar_cross_orbit}

\begin{figure*}
    \centering
    \includegraphics[width=.85\textwidth]{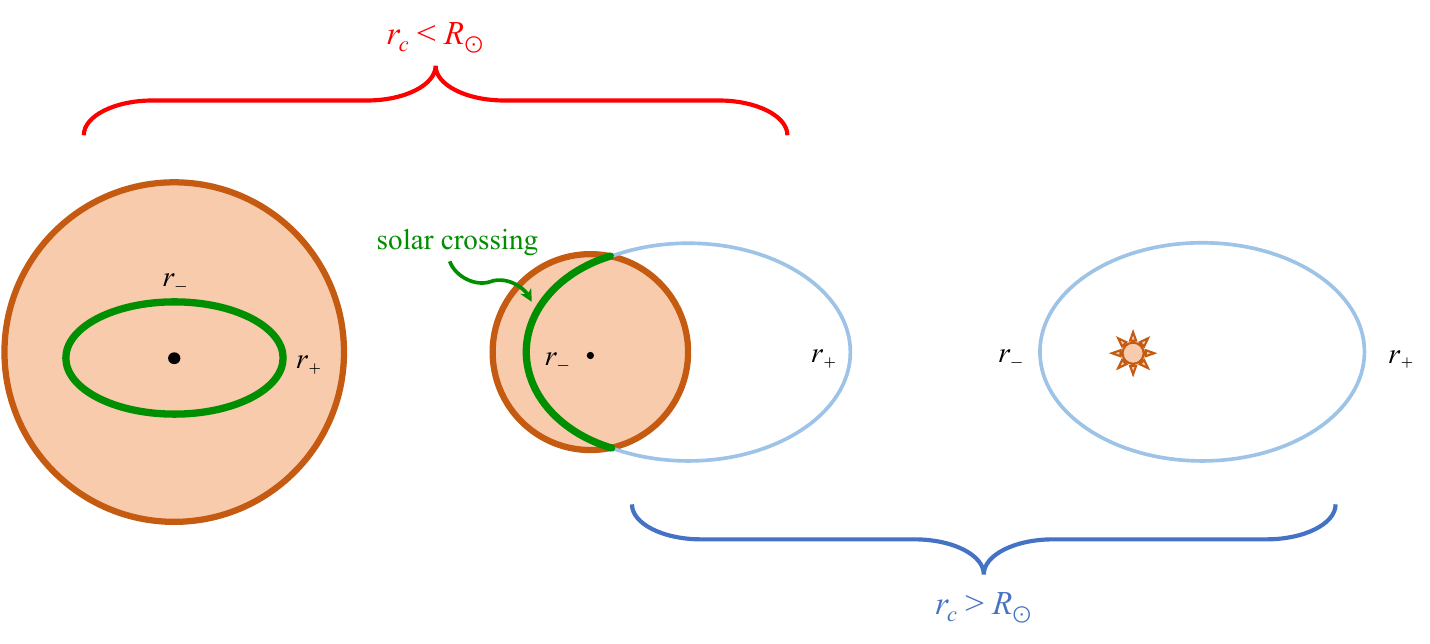}
    \caption{Bound-orbit classes relative to the Sun (orange).
    From left to right: an interior orbit with $r_+<R_\odot$, a
    solar-crossing orbit with $r_-<R_\odot<r_+$, and an exterior
    orbit with $r_->R_\odot$.
    Green and light-blue segments lie inside and outside the Sun,
    respectively. $r_-$ and $r_+$ mark the pericenter and apocenter.
    Only in the first two cases can DM scatter with solar nuclei.
    }
    \label{fig:orbit_types}
\end{figure*}

A bound particle can scatter on solar nuclei only if its orbit enters the Sun, $r_-<R_\odot$.
Fig.~\ref{fig:orbit_types} shows the three classes of bound orbits, namely orbits inside the Sun (left), orbits crossing its surface (middle) and orbits lying outside it (right).
Particles that remain bound after capture or scattering inside the Sun move on orbits of the first two classes and make up the population considered in this paper.

\begin{figure}
    \centering
    \includegraphics[width=.9\columnwidth]{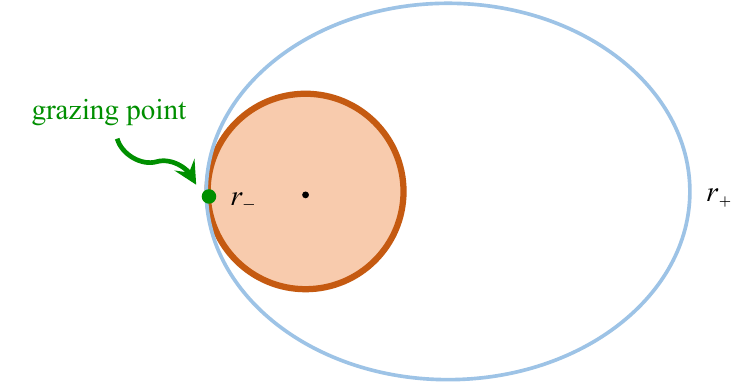}
    \caption{An orbit with a grazing point, $r_c>R_\odot$ and
    $L=L_\odot(E)<L_c(E)$, will not cross the solar interior.
    }

\label{fig:graze_case}
\end{figure}

At fixed energy, increasing $L$ raises $r_-$ and lowers $r_+$ until they coincide on a circular orbit, $r_-=r_+=r_c(E)$.
The angular momentum of this circular orbit, $L_c(E)$, is the largest allowed at that energy.
If $r_c(E)<R_\odot$, every orbit at that energy has $r_-<R_\odot$ and crosses the solar interior, although its $r_+$ may lie outside the Sun.
If $r_c(E)>R_\odot$, decreasing $L$ moves $r_-$ inward until it reaches the solar surface at $L=L_\odot(E)$, as shown in
Fig.~\ref{fig:graze_case}.
At this grazing point the velocity is purely tangential, so $L_\odot(E)=R_\odot v$ with $E=v^2/2+\Phi(R_\odot)$.
Including the case $r_c(E)\leq R_\odot$, we obtain
\begin{equation}
  L_\odot(E)=
  \begin{cases}
    L_c(E),&r_c(E)\leq R_\odot,\\
    R_\odot\sqrt{2[E-\Phi(R_\odot)]},&r_c(E)>R_\odot.
  \end{cases}
  \label{eq:solar-angular-boundary}
\end{equation}
In the second case, $L_\odot<L_c$, and orbits with $L<L_\odot$ pass through the Sun, whereas those with $L_\odot<L\leq L_c$ remain outside it.

\section{Derivations in orbital phase space}\label{app:bte_in_xi}

In this appendix, we collect the steps that take the BTE from position and velocity space to the orbital phase space of the main text, for the static, spherically symmetric potential assumed there.
Each phase-space point $\rv$ lies on a single orbit $\xi=(E,L)$, so integrals over $\rv$ can be organized orbit by orbit.
For later use, primed variables denote a phase-space point $(\bfr',\bfv')$ with orbital coordinates $\xi'\equiv\xi(\bfr',\bfv')$.

\subsection{The vanishing advection terms}\label{app:0_advection}

Multiplying the advection terms of Eq.~\eqref{eq:boltzmann} by $\Delta(\xi-\xi')$ and integrating over phase space, we obtain
\begin{align}
  &\int d^3\bfr'd^3\bfv'\,\Delta(\xi-\xi')(\bfv'\cdot\nabla_{\bfr'}f-\nabla_{\bfr'}\Phi\cdot\nabla_{\bfv'}f)\nonumber\\
  &=-\int d^3\bfr'd^3\bfv'\,f\,(\bfv'\cdot\nabla_{\bfr'}-\nabla_{\bfr'}\Phi\cdot\nabla_{\bfv'})\Delta(\xi-\xi')=0.
  \label{eq:advection_projection}
\end{align}
The first equality follows from integration by parts with $\nabla_{\bfr'}\cdot\bfv'=0$ and $\nabla_{\bfv'}\cdot\nabla_{\bfr'}\Phi=0$, and the boundary terms vanish because the collisionless motion conserves $E$ and does not leave the bound domain.
The second equality holds because $E'$ and $L'$ are constant along the collisionless motion, so the operator in parentheses gives zero when acting on any function of $(E',L')$.
The projection therefore removes the advection terms without any assumption on $f(\bfr',\bfv')$.

\subsection{From \texorpdfstring{$(\bfr,\bfv)$}{(r,v)} to orbital phase space}
\label{app:rv_to_EL}

At a local phase-space point $(\bfr',\bfv')$, the specific energy and angular momentum are
\begin{equation}
E' = \frac{v'^2}{2} + \Phi(r'),\quad L'=|\bfr'\times \bfv'| \label{eq:E'_and_L'}
\end{equation}
with
$d^3\bfv'=v_t' dv_t'dv_r'd\alpha$.
After integrating over the spatial orientation and $\alpha$, 
we obtain
\begin{align}
 &\Delta(\xi-\xi') d^3\bfr'd^3\bfv' \nonumber \\
 &= 2\times 8\pi^2 r'^2 v_t' \Delta(\xi-\xi') dr'dv_r'dv_t'.  \label{eq:d3r'd3v'}
\end{align}
The factor of two accounts for the inward and outward passages, which contribute equally, while the angular integrations
give $4\pi$ from the direction of $\bfr'$ and $2\pi$ from the azimuth $\alpha$ of $\bfv'$ about it.
The delta functions in $\Delta(\xi-\xi')$ fix the orbital coordinates to $\xi$, and
the remaining Jacobian is
\begin{equation}
    dv_r'dv_t' = \left|\frac{\partial(v_r',v_t')}{\partial(E',L')}\right|dE'dL' = \frac{dE' dL'}{r' v_r'}\label{eq:jac_v_2_EL}
\end{equation}
such that
\begin{equation*}
     16\pi^2 r'^2 v_t' dr'dv_r'dv_t' = 16\pi^2 L' \frac{dr'}{v_r'(r',\xi')}dE'dL'.
\end{equation*}
We then obtain
\begin{align}
    &\Delta(\xi-\xi') d^3 \bfr' d^3\bfv' \nonumber \\
    &=   16\pi^2 L' \frac{dr'}{v_r'(r',\xi')}\Delta(\xi-\xi') dE'dL'\label{eq:spatial_to_xi}
\end{align}
and
\begin{align}
    dV_\xi &\equiv dEdL \int  d^3\bfr'd^3\bfv'\,\Delta(\xi-\xi')  \nonumber\\
     &= dEdL \int 16\pi^2 L' \frac{dr'}{v_r'(r',\xi')}\Delta(\xi-\xi') dE'dL' \nonumber \\
     & = 8\pi^2 L\tau_r(\xi) dEdL
\end{align}
is the volume element of orbital phase space.
Weighting Eq.~\eqref{eq:spatial_to_xi} by the orbit-averaged distribution gives
\begin{align}
    & dEdL \int d^3\bfr'd^3\bfv'\,\Delta(\xi-\xi') f(\bfr',\bfv') \nonumber \\
    & = dEdL  \int 16\pi^2  L'\frac{dr'}{v_r'(r',\xi')}\Delta(\xi-\xi') f(\xi')  dE'dL' \nonumber \\
    & = 8\pi^2 L\tau_r(\xi)f(\xi) dEdL \equiv n_\chi(\xi)dEdL, \label{eq:from_spatial_f_to_xi_f}
\end{align}
where $f(\xi)$ is the distribution averaged over the orbit, that is, over its orientations and over the time spent at each radius.
Since the integral covers the whole orbit, replacing $f$ by its orbit average in the second line does not require any approximation.
The orbit-averaging approximation is needed only when $f$ multiplies a quantity that varies along the orbit, as in the collisional kernel of App.~\ref{app:collisional_kernel}.
The number of particles in an orbital interval is then
\begin{equation}
    dN_\chi = n_\chi(\xi) dEdL = f(\xi)dV_\xi \label{eq:dNx_xi_app}.
\end{equation}
Equivalently,
\begin{equation}
    f(\xi) = \frac{dN_\chi}{dV_\xi} = \frac{\int d^3\bfr'd^3\bfv'\,\Delta(\xi-\xi')f(\bfr',\bfv')}{8\pi^2 L\tau_r(\xi)}\label{eq:f(xi)}
\end{equation}
relates the local distribution to its orbital average.

\subsection{Derivation of the master equation}\label{app:master_derivation}

The phase-space volume of a fixed orbit is time independent.
Differentiating Eq.~\eqref{eq:f(xi)} with respect to $t$, substituting Eq.~\eqref{eq:boltzmann} and using Eq.~\eqref{eq:advection_projection} gives
\begin{align*}
    &\int d^3\bfr'd^3\bfv'\,\Delta(\xi-\xi') \frac{\partial f(\bfr',\bfv')}{\partial t} \nonumber\\
    &=
    \int d^3\bfr'd^3\bfv'\,\Delta(\xi-\xi') \mathcal{C}[f](\bfr',\bfv').
\end{align*}
Dividing by the orbital volume factor gives
\begin{equation}
\frac{\partial f(\xi)}{\partial t} = \frac{\int d^3\bfr'd^3\bfv'\,\Delta(\xi-\xi') \mathcal{C}[f](\bfr',\bfv')}{8\pi^2 L \tau_r(\xi)}.
\end{equation}
Multiplying by $8\pi^2L\tau_r(\xi)$ and using Eq.~\eqref{eq:nx_xi} then yields
\begin{equation}
 \frac{\partial n_\chi(\xi)}{\partial t}=
 \int d^3\bfr'd^3\bfv'\,\Delta(\xi-\xi') \mathcal{C}[f](\bfr',\bfv')
\end{equation}
which is the master equation, Eq.~\eqref{eq:master_eq}.

\begin{widetext}
\subsection{Derivation of the collisional kernel}
\label{app:collisional_kernel}

Substituting Eq.~\eqref{eq:Ccoll_f} into the
master equation, Eq.~\eqref{eq:master_eq},  gives
the gain and loss terms below
\begin{align}
    &\int d^3\bfr'd^3\bfv'\,\Delta(\xi-\xi') \Ccoll[f](\bfr',\bfv') \nonumber \\
    &= \int d^3\bfr'd^3\bfv'\,\Delta(\xi-\xi') \int d^3\bfv''\,[R(\bfv''\to\bfv',r')f(\bfr',{\bf v}'')
  -R(\bfv'\to\bfv'',r')f(\bfr',\bfv')].
\end{align}
For the loss term, Eq.~\eqref{eq:spatial_to_xi} and the orbit-averaging approximation, Eq.~\eqref{eq:orbit_average}, give
\begin{align}
&\int d^3\bfr'd^3\bfv'\,\Delta(\xi-\xi')f(\bfr',\bfv') \int d^3\bfv''R(\bfv'\to\bfv'',r') \nonumber \\
&= \int 16\pi^2 L' \frac{dr'}{v_r'(r',\xi')}\Delta(\xi-\xi') dE'dL'\,f(\xi')\int d^3\bfv''R(\bfv'\to\bfv'',r').\label{eq:loss_kernel_middle}
\end{align}
After integrating over the outgoing velocity orientation, the delta function in $R(\bfv'\to \bfv'')$ gives
\begin{equation}
d^3\bfv''\,\delta^{(3)}(\bfv''-\bfv''_f)=dE''dL''\,\delta(E''-E''_f)\delta(L''-L''_f)\label{eq:d3v_delta3}
\end{equation}
which is proved in App.~\ref{app:identity_d3v_delta3}.
The final velocity $\bfv_f''$ fixes $(E_f'',L_f'')$ through the kinematics
given in App.~\ref{app:kinematics}.
Therefore
\begin{align}
\int d^3\bfv''R(\bfv'\to\bfv'',r') &=\int dE''dL'' \sum_A \int d\Omega_u\int_0^\infty du\,u^2f_A(r',u)n_A(r')\nonumber\\
&\qquad\times
  \frac{\sigma_{\chi A}}{4\pi}\int d\Omega_{g'}gF_A^2(q) \delta(E''-E''_f)\delta(L''-L''_f)
\end{align}
Substituting into Eq.~\eqref{eq:loss_kernel_middle} and integrating over $dE'dL'$ yields
\begin{align}
    16\pi^2 L \int_{r_-(\xi)}^{r_+(\xi)} \frac{dr'}{v_r(r',\xi)}&f(\xi)\int dE''dL'' \sum_A n_A(r')\int d\Omega_u\int_0^\infty du\,u^2f_A(r',u)
     \nonumber \\
    & \times
  \frac{\sigma_{\chi A}}{4\pi}\int d\Omega_{g'}gF_A^2(q)\delta(E''-E''_f)\delta(L''-L''_f).
\end{align}
Relabeling the integration variables $(r',E'',L'',E''_f,L''_f)\to(r,E',L',E_f',L_f')$ gives
\begin{align}
   \int dE'dL' & n_\chi(\xi) \times\left[ \frac{2}{\tau_r(\xi)} \int_{r_-(\xi)}^{r_+(\xi)} \frac{dr}{v_r(r,\xi)}\right.
     \nonumber \\
    & \left.\times \sum_A n_A(r)\int d\Omega_u\int_0^\infty du\,u^2f_A(r,u)
  \frac{\sigma_{\chi A}}{4\pi}\int d\Omega_{g'}gF_A^2(q)\delta(E'-E_f')\delta(L'-L_f')\right] \nonumber \\
  & = \int_{\Phi(0)}^{0} dE' \int_0^{L_\odot(E')}dL'\,K(\xi\to\xi')n_\chi(\xi).\label{eq:loss_term}
\end{align}
The expression in brackets is $K(\xi\to\xi')$ in Eq.~\eqref{eq:rate_K} and
the integration limits select bound destinations.
For the gain term,
\begin{align}
&\int d^3\bfr'd^3\bfv'\,\Delta(\xi-\xi') \int d^3\bfv''R(\bfv''\to\bfv',r')f(\bfr',\bfv'') \nonumber \\
&=  \int d^3\bfr'd^3\bfv''\,f(\bfr',\bfv'') \int d^3\bfv'R(\bfv''\to\bfv',r')\Delta(\xi-\xi') \nonumber \\
&= \int 16\pi^2 L'' \frac{dr'}{v_r(r',\xi'')}f(\xi'')dE''dL'' \int d^3\bfv'R(\bfv''\to\bfv',r')\Delta(\xi-\xi') \nonumber \\
& =\int dE''dL''\,n_\chi(\xi'')  \times \left[\frac{2}{\tau_r(\xi'')}\int_{r_-(\xi'')}^{r_+(\xi'')} \frac{dr'}{v_r(r',\xi'')}\int d^3\bfv'R(\bfv''\to\bfv',r')\Delta(\xi-\xi')\right].\label{eq:gain_term_middle}
\end{align}
The third line uses Eq.~\eqref{eq:spatial_to_xi} for the incident state $\xi''$ and the orbit-averaging approximation, $f(\bfr',\bfv'')\simeq f(\xi'')$.
Applying the identity in Eq.~\eqref{eq:d3v_delta3}, we obtain
\begin{align}
    &\int d^3\bfv'R(\bfv''\to\bfv',r')\Delta(\xi-\xi') \nonumber \\
    &=
    \int dE'dL'  \sum_A n_A(r')\int d\Omega_u\int_0^\infty du\,u^2f_A(r',u)
  \frac{\sigma_{\chi A}}{4\pi}\int d\Omega_{g'}gF_A^2(q) \delta(E'-E'_f)\delta(L'-L'_f)\Delta(\xi-\xi') \nonumber \\
  & =\sum_A n_A(r')\int d\Omega_u\int_0^\infty du\,u^2f_A(r',u)
  \frac{\sigma_{\chi A}}{4\pi}\int d\Omega_{g'}gF_A^2(q) \delta(E-E'_f)\delta(L-L'_f).
\end{align}
Substituting into Eq.~\eqref{eq:gain_term_middle} and relabeling
$(r',E'',L'',\xi'',E_f',L_f')\to(r,E',L',\xi',E_f,L_f)$ gives
\begin{align}
    \int dE'dL' &n_\chi(\xi') \times \left[ \frac{2}{\tau_r(\xi')}
    \int_{r_-(\xi')}^{r_+(\xi')}\frac{dr}{v_r(r,\xi')}\right. \nonumber \\
    &\left.\times \sum_A n_A(r)\int d\Omega_u\int_0^\infty du\,u^2f_A(r,u)
  \frac{\sigma_{\chi A}}{4\pi}\int d\Omega_{g'}gF_A^2(q) \delta(E-E_f)\delta(L-L_f)\right]\nonumber \\
  & =\int_{\Phi(0)}^{0} dE' \int_0^{L_\odot(E')}dL'\,K(\xi'\to\xi)n_\chi(\xi').\label{eq:gain_term}
\end{align}
Combining Eqs.~\eqref{eq:loss_term} and \eqref{eq:gain_term}, we obtain the collisional kernel
\begin{equation}
\Kcoll[n_\chi](\xi)= \int_{\Phi(0)}^{0} dE' \int_0^{L_\odot(E')}dL'\,[K(\xi'\to \xi)n_\chi(\xi')-K(\xi\to\xi')n_\chi(\xi)]
\end{equation}
\end{widetext}
which reproduces Eq.~\eqref{eq:Kcoll}, and the evaporation and annihilation kernels follow from the same projection.

\begin{figure}
 \centering
 \includegraphics[width=.48\textwidth]{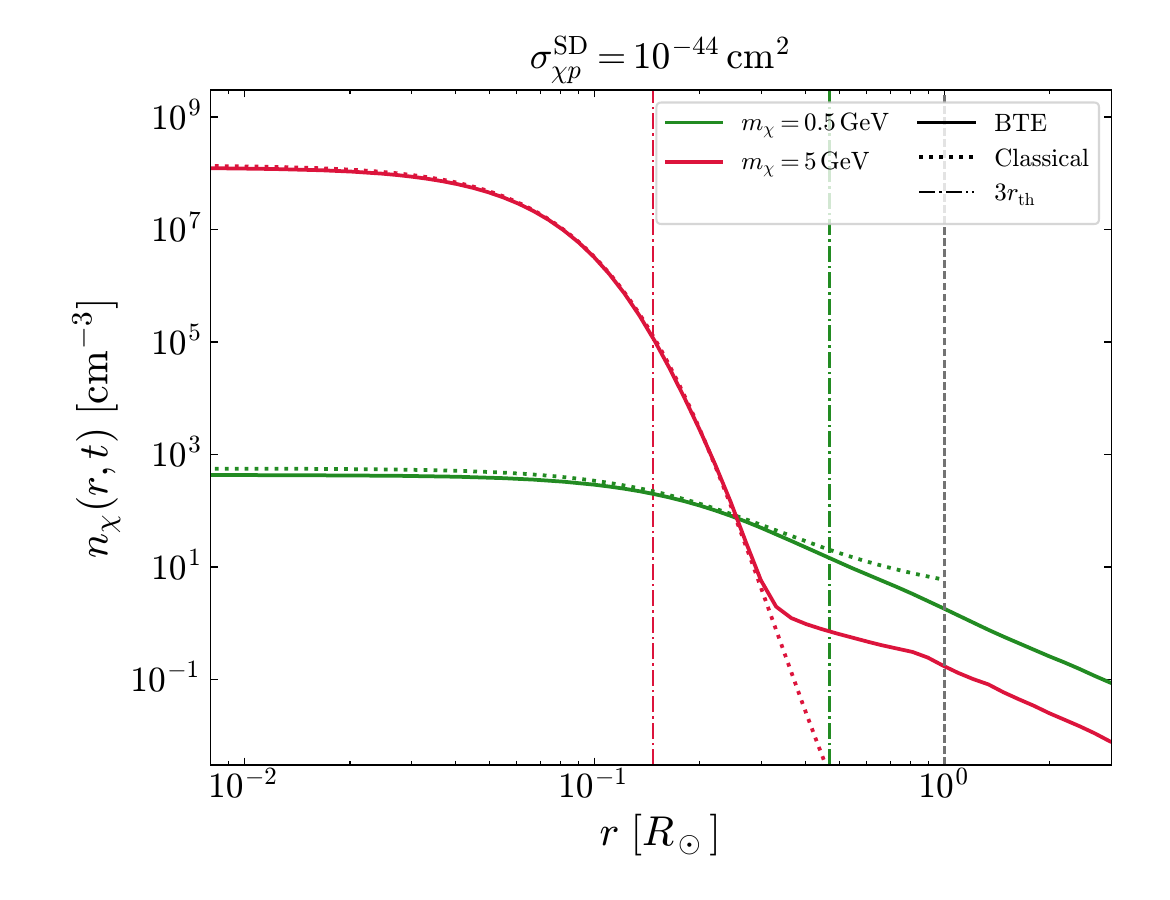}
 \caption{$n_\chi(r,t)$
 at $t_\odot$ for SD scattering with
 $\sigma_{\chi p}=10^{-44}~{\rm cm^2}$ and DM masses labeled 0.5~GeV (green)
 and 5~GeV (red). Solid and dotted curves show the BTE and classical results.
 Dot-dashed lines mark $3r_{\rm th}$ for each mass and the gray dashed line marks
 the solar surface.}

\label{fig:nx_tail}
\end{figure}

\subsection{Proof of Eq.~\eqref{eq:d3v_delta3}}
\label{app:identity_d3v_delta3}

For a test function $h$, the defining property of the delta function is
\begin{equation}
\int d^3\bfv\,\delta^{(3)}(\bfv -\bfv_0) h(\bfv) = h(\bfv_0)
\end{equation}
Using $d^3\bfv=v_tdv_tdv_rd\alpha$, this becomes
\begin{equation*}
\int dv_tdv_rd\alpha\,v_t\delta^{(3)}(\bfv -\bfv_0)  h(v_t,v_r, \alpha)=h(v_{t,0},v_{r,0}, \alpha_0)
\end{equation*}
where the right-hand side can also be written as
\begin{align*}
&h(v_{t,0},v_{r,0}, \alpha_0) \nonumber \\
&= \int dv_tdv_rd\alpha\,\delta(v_t-v_{t,0})
\delta(v_r-v_{r,0})\delta(\alpha-\alpha_0) \nonumber \\
&\qquad \times h(v_t,v_r,\alpha).
\end{align*}
Comparing these expressions gives
\begin{align}
&\delta^{(3)}(\bfv -\bfv_0)\nonumber \\
&= \frac{1}{v_t}\delta(v_t-v_{t,0})
\delta(v_r-v_{r,0})\delta(\alpha-\alpha_0).
\end{align}
From Eq.~\eqref{eq:jac_v_2_EL}, we have
$d^3{\bf v}=v_t dEdLd\alpha/(r v_r)$, giving
\begin{align}
&d^3{\bf v}\delta^{(3)}(\bfv -\bfv_0) \nonumber \\
&= \frac{dEdLd\alpha}{r v_r} \delta(v_t-v_{t,0})
\delta(v_r-v_{r,0})\delta(\alpha-\alpha_0).
\end{align}
We also have
\begin{equation*}
 \delta(v_t-v_{t,0})=\delta\left(\frac{L}{r} - \frac{L_0}{r}\right)
 = r\delta(L-L_0)
\end{equation*}
and
\begin{equation*}
E-E_0 = \frac{(v_r^2+v_t^2)-(v_{r,0}^2+v_{t,0}^2)}{2}=\frac{v_r^2-v_{r,0}^2}{2},
\end{equation*}
where we have used the constraint $v_t=v_{t,0}$ from $\delta(v_t-v_{t,0})$. The delta function selects the radial direction of \(\bfv_0\), so
\begin{equation*}
\delta(E-E_0) = \delta\left(\frac{v_r^2-v_{r,0}^2}{2}\right)=\frac{\delta(v_r-v_{r,0})}{v_r}.
\end{equation*}
Finally, we obtain
\begin{align}
d^3{\bfv}\delta^{(3)}(\bfv-\bfv_0) &=
     \delta(E-E_0)
\delta(L-L_0)dEdL \nonumber \\ 
&\qquad \times \int \delta(\alpha-\alpha_0)d\alpha \nonumber \\
 &= \delta(E-E_0)
\delta(L-L_0)dEdL
\end{align}
which is Eq.~\eqref{eq:d3v_delta3} with the direction of velocity
integrated out.

\section{$n_\chi(r)$ tail at $r>r_{\rm th}$}
\label{app:nx_tail}

\begin{figure}
    \centering
    \includegraphics[width=.48\textwidth]{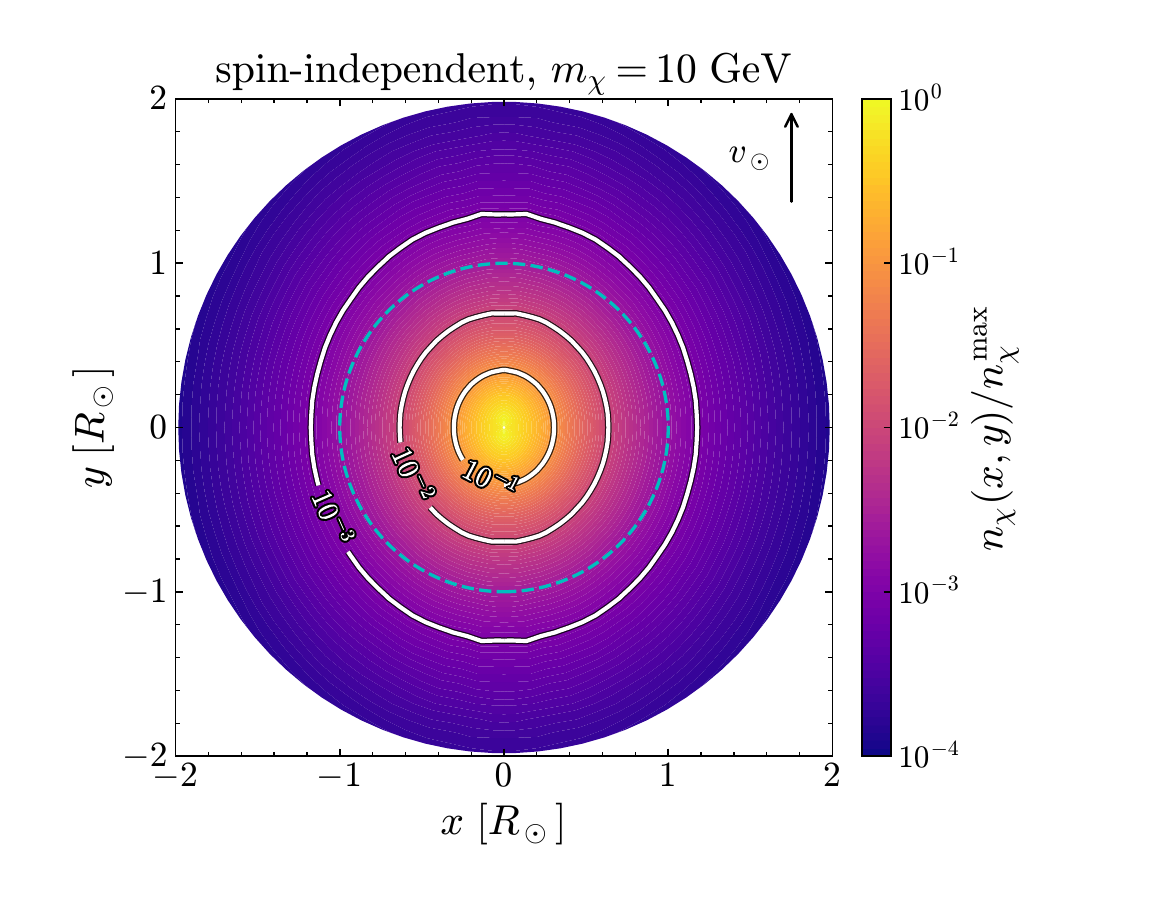}
    \caption{Density of initially captured DM for SI scattering at $m_\chi=10$~GeV without subsequent collisions, normalized to its maximum. The black arrow indicates the motion of the Sun, the dashed circle marks the solar surface, and the contours mark $10^{-1}$, $10^{-2}$ and $10^{-3}$.}
    \label{fig:anisotropy}
\end{figure}

Fig.~\ref{fig:nx_tail} illustrates the $n_\chi$ tail responsible for the evaporation
behavior discussed in Sec.~\ref{sec:evap_mass}.
For both masses, the BTE and classical radial densities agree closely within
$3r_{\rm th}$. At $m_\chi=0.5~{\rm GeV}$ they remain comparable out to
the solar surface, consistent with the modest change in the evaporation
coefficient ratio between the two orbital selections in Fig.~\ref{fig:C_e_ratio}.
At $m_\chi=5~{\rm GeV}$, the profiles separate beyond roughly $0.3R_\odot$, where freshly captured, weakly bound particles sustain a tail that can still evaporate while the classical density is exponentially suppressed, which explains the large coefficient ratio in Fig.~\ref{fig:C_e_ratio}.

\section{Anisotropy of rarely scattering DM}
\label{app:anisotropy}
\begin{figure}
    \centering
    \includegraphics[width=.48\textwidth]{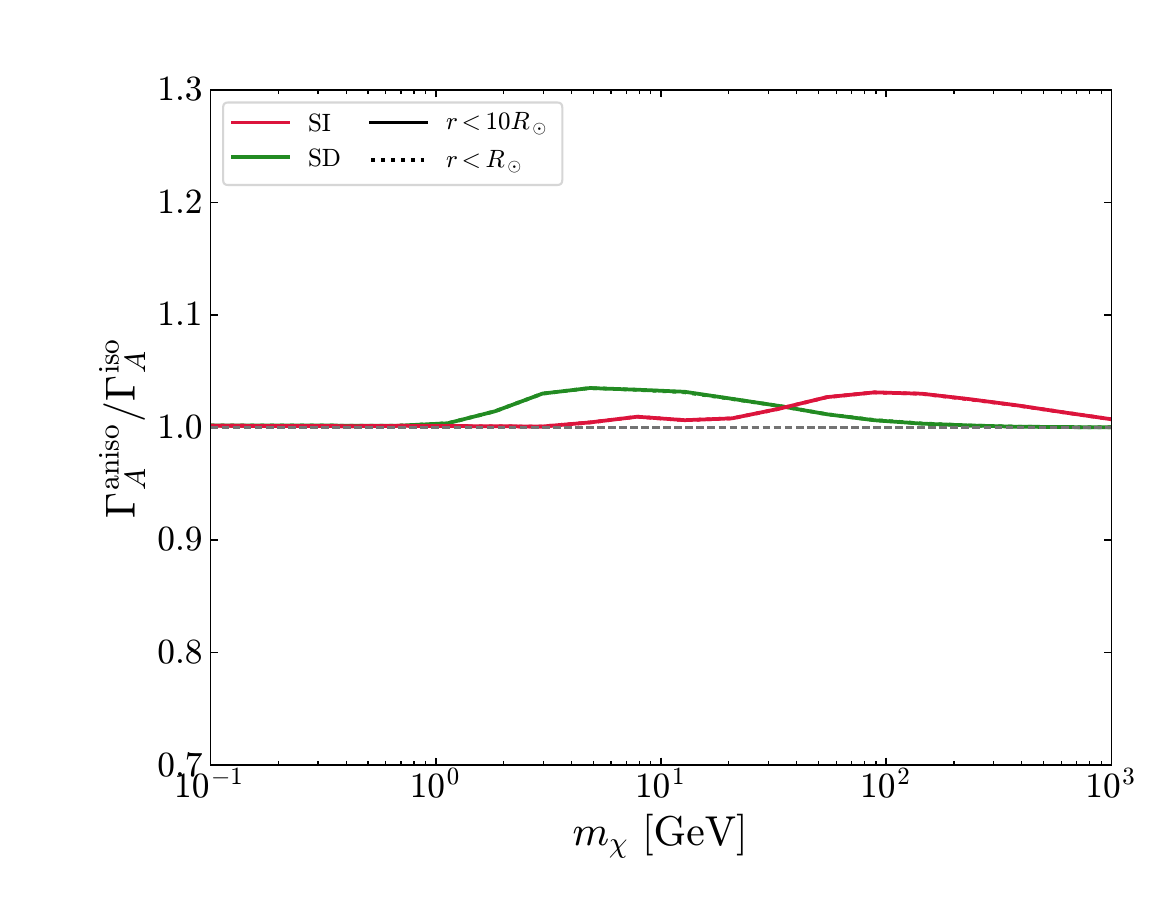}
    \caption{$\Gamma_A^{\rm aniso}/\Gamma_A^{\rm iso}$ for SI and SD scattering. Both rates are integrated within $10R_\odot$ (solid) and $R_\odot$ (dotted).}
    \label{fig:aniso_ratio}
\end{figure}

When collisions are rare, the orbital planes of captured DM keep a memory of the incident directions of halo DM.
To estimate the largest possible effect, we assume that the initial anisotropy is never erased.
We sample the positions and velocities of captured DM with a Monte Carlo simulation of the initial capture, based on the scattering kinematics of App.~\ref{app:kinematics}.
From this population, we compute the annihilation rate $\Gamma_A^{\rm aniso}$ and compare it with $\Gamma_A^{\rm iso}$ obtained from the spherically symmetric density of Eq.~\eqref{eq:nx(r)_reconstruction}.
Fig.~\ref{fig:anisotropy} shows the density for SI scattering at $m_\chi=10$~GeV, which is elongated along the solar motion.
Fig.~\ref{fig:aniso_ratio} shows that $\Gamma_A^{\rm aniso}/\Gamma_A^{\rm iso}$ stays within 4\% of unity for all masses considered in this paper, and SD scattering gives the same result.

\section{Grid in orbital phase space}
\label{app:grid}
We evolve $n_\chi(\xi)$ on 192 cells in $E$ and 48 cells in $\ell\equiv L/L_\odot(E)$, where $L_\odot(E)$ is given by Eq.~\eqref{eq:solar-angular-boundary}, so that $\ell=0$ for radial orbits and $\ell=1$ at the largest angular momentum.
The grid depends only on $m_\chi$,
and all cross sections share it.
Once the grid is set, we integrate Eq.~\eqref{eq:rate_K} over the cells to obtain the rate of transfer between every pair of cells, which redistributes the bound population through Eq.~\eqref{eq:Kcoll}, and the evaporation rate of each cell.
Since these rates and the capture rate of each cell, obtained from Eq.~\eqref{eq:orbital-capture-source}, are proportional to $\sigma_{\chi p}$, we compute them once for each mass and type of scattering, and the master equation becomes a set of coupled equations for the cell populations.

The energy range runs from the bottom of the potential, $\Phi(0)$, to orbits whose radial turning point lies near $1100R_\odot$, close to the orbit of Jupiter, and we split it into three regions.
In terms of $\epsilon\equiv E-\Phi(0)$, the first region covers $0\leq\epsilon\leq10\,\epsilon_0$ with 120 cells, where $\epsilon_0\equiv T_\odot(0)/m_\chi$ is the energy scale of DM thermalized with the solar core.
Their edges are equally spaced in $\ln(\epsilon+\epsilon_0)$, so the cells are nearly uniform for $\epsilon\ll\epsilon_0$ and widen in proportion to $\epsilon$ above it.
Since $\epsilon_0$ scales as $1/m_\chi$, thermalized DM occupies a thinner layer for heavier DM, and we rebuild this region for each mass so that the same number of cells always resolves the thermal core.
The second region contains 60 cells whose widths grow by a constant factor up to a turning radius near $12R_\odot$, which covers DM that cools after capture.
The third region contains 12 cells with mass-independent edges between $12R_\odot$ and $1100R_\odot$, where newly captured DM resides.
Bound DM on wider orbits beyond this range is kept in a reservoir, which counts as bound but does not scatter or annihilate.
For $m_\chi\lesssim3$~GeV, $10\,\epsilon_0$ is comparable to the depth of the potential, and the first region ends at a lower energy so that the cells of the second region have equal widths.

We
space the 48 cells in $\ell$ uniformly instead of logarithmically, because bound DM spreads over the whole range of $\ell$ and $dN_\chi/d\ell$ vanishes as $\ell\to0$, so logarithmic cells would crowd the sparsely populated radial orbits.
We use this grid because a finer energy grid, at a higher computational cost, changes $\Gamma_A$ by at most a factor of $\mathcal{O}(1)$, which is small compared with the suppression in Figs.~\ref{fig:Gamma_A_ratio} and \ref{fig:Gamma_A_ratio_scan}.

\bibliographystyle{apsrev4-2-inspire}
\bibliography{main}

\end{document}